\documentstyle[amstex,amssymb,12pt]{article}
\begin{document}

\author{{\bf Fabio Cardone}$^{a,b}${\bf , Alessio Marrani}$^{c,d}${\bf \ and } \and
{\bf Roberto Mignani}$^{b-d}$ \\
$a$ Dipartimento di Fisica\\
Universit\`{a} dell'Aquila\\
Via Vetoio \\
67010 Coppito, L'Aquila, Italy\\
$b$ I.N.D.A.M. - G.N.F.M.\\
$c$ Universit\`{a} degli Studi ''Roma Tre''\\
Via della Vasca Navale, 84\\
00146 ROMA, Italy\\
$d$ I.N.F.N. - Sezione di Roma III}
\title{{\bf Killing symmetries of generalized Minkowski spaces.}\\
{\bf 1- Algebraic-infinitesimal structure of space-time rotation groups}}
\maketitle

\begin{abstract}
In this paper, we introduce the concept of N-dimensional generalized
Minkowski space, i.e. a space endowed with a (in general non-diagonal)
metric tensor, whose coefficients do depend on a set of non-metrical
coodinates. This is the first of a series of papers devoted to the
investigation of the Killing symmetries of generalized Minkowski spaces. In
particular, we discuss here the infinitesimal-algebraic structure of the
space-time rotations in such spaces. It is shown that the maximal Killing
group of these spaces is the direct product of a generalized Lorentz group
and a generalized translation group. We derive the explicit form of the
generators of the generalized Lorentz group in the self-representation and
their related, generalized Lorentz algebra. The results obtained are
specialized to the case of a 4-dimensional, ''deformed'' Minkowski space $%
\widetilde{M_{4}}$, i.e. a pseudoeuclidean space with metric coefficients
depending on energy.
\end{abstract}

\bigskip \bigskip

\bigskip

\section{Introduction\protect\bigskip}

In the last years, two of the present authors (F.C. and R.M.) proposed a
generalization\medskip\ of {\em Standard Special Relativity }(SR) based on
a\medskip\ ''deformation'' of space-time, assumed to be endowed with a
metric whose coefficients depend on the energy of\medskip\ the process
considered [1]. Such a formalism ({\em Deformed Special Relativity}, DSR)
applies in principle to {\em all\medskip } four interactions
(electromagnetic, weak, strong and gravitational) - at least as far as their
nonlocal behavior and\medskip\ nonpotential part is concerned - and provides
a metric representation of them (at\medskip\ least for the process and in
the\medskip\ energy range considered) ([1]-[5]). Moreover, it was shown that
such a formalism is actually a\medskip\ five-dimensional one, in
the\medskip\ sense that the deformed Minkowski space is embedded in\medskip\
a larger Riemannian manifold, with energy as fifth dimension [6].\medskip

\bigskip In this paper, following the line of mathematical-formal research
started with [7] and [8], we introduce the concept of $N$-dimensional
generalized Minkowski space, i.e. a space endowed with a (in general
non-diagonal) metric tensor, whose coefficients do depend on a set of
non-metrical coodinates. The deformed space-time $\widetilde{M_{4}}$ of DSR
is just a special case of such spaces. This is the first of a series of
papers devoted to the investigation of the Killing symmetries of generalized
Minkowski spaces. In particular, we shall discuss here the
infinitesimal-algebraic structure of the space-time rotations in such spaces.

The organization of the paper is as follows. In Sect. 2 we briefly review
the formalism of DSR4 \ and of the deformed Minkowski space $\widetilde{M_{4}%
}$. Generalized Minkowski spaces are defined in Subsect. 3.1. In Subsect.
3.2 we look for the group of isometries of such spaces by means of the
Killing equations. It is shown that the maximal Killing group of these
spaces is the semidirect product of a generalized Lorentz group and a
generalized translation group. The infinitesimal structure of the
generalized Lorentz group is discussed in Sect. 4, where we derive the
explicit form of its generators in the self-representation and their
related, generalized Lorentz algebra. The special case of the deformed space
$\widetilde{M_{4}}$ of DSR4 is considered in Sect. 5. Sect. 6 concludes the
paper.

\bigskip

\section{\protect\bigskip Deformed Special Relativity in four dimensions
(DSR4)}

The generalized (``deformed'') Minkowski space $\widetilde{M_{4}}$ (DMS4) is
defined as a space with the same local coordinates $x$ of $M_{4}$ (the
four-vectors of the usual Minkowski space), but with metric given by the
metric tensor\footnote{%
In the following, we shall employ the notation ''ESC on'' (''ESC off'') to
mean that the Einstein sum convention on repeated indices is (is not) used.}
\begin{gather}
g_{\mu \nu
,DSR4}(x^{5})=diag(b_{0}^{2}(x^{5}),-b_{1}^{2}(x^{5}),-b_{2}^{2}(x^{5}),-b_{3}^{2}(x^{5}))=
\nonumber \\
\nonumber \\
\stackrel{\text{{\footnotesize ESC off}}}{=}\delta _{\mu \nu }\left[
b_{0}^{2}(x^{5})\delta _{\mu 0}-b_{1}^{2}(x^{5})\delta _{\mu
1}-b_{2}^{2}(x^{5})\delta _{\mu 2}-b_{3}^{2}(x^{5})\delta _{\mu 3}\right]
\end{gather}
where the $\left\{ b_{\mu }^{2}(x^{5})\right\} $ are dimensionless, real,
positive functions of the independent, non-metrical (n.m.) variable $x^{5}$
\footnote{%
Such a coordinate is to be interpreted as the energy (see Refs. [1]-[5]);
moreover, the index $5$ explicitly refers to the above-mentioned fact that
the deformed Minkowski space can be {\em ''naturally'' embedded} in a
five-dimensional (Riemannian) space [6].}. The generalized interval in $%
\widetilde{M_{4}}$ is therefore given by ($x^{\mu
}=(x^{0},x^{1},x^{2},x^{3})=(ct,x,y,z)$, with $c$ being the usual light
speed in vacuum):
\begin{equation}
ds^{2}=b_{0}^{2}c^{2}\left( dt\right) ^{2}-\left[ b_{1}^{2}\left( dx\right)
^{2}+b_{2}^{2}\left( dy\right) ^{2}+b_{3}^{2}\left( dz\right) ^{2}\right]
=g_{\mu \nu ,DSR4}dx^{\mu }dx^{\nu }=dx\ast dx.
\end{equation}

The last step in (2) defines the scalar product $\ast $ in the deformed
Minkowski space $\widetilde{M_{4}}$ .\bigskip\ In order to emphasize the
dependence of DMS4 on the variable $x^{5}$, we shall sometimes use the
notation $\widetilde{M_{4}}(x^{5})$. It follows immediately that it can be
regarded as a particular case of a Riemann space with null curvature.

From the general condition
\begin{equation}
g_{\mu \nu ,DSR4}(x^{5})g_{DSR4}^{\nu \rho }(x^{5})=\delta _{\mu }^{~~\rho }
\end{equation}
we get for the contravariant components of the metric tensor
\begin{gather}
g_{DSR4}^{\mu \nu
}(x^{5})=diag(b_{0}^{-2}(x^{5}),-b_{1}^{-2}(x^{5}),-b_{2}^{-2}(x^{5}),-b_{3}^{-2}(x^{5}))=\medskip
\nonumber \\
\nonumber \\
\stackrel{\text{{\footnotesize ESC off}}}{=}\delta ^{\mu \nu }\left(
b_{0}^{-2}(x^{5})\delta ^{\mu 0}-b_{1}^{-2}(x^{5})\delta ^{\mu
1}-b_{2}^{-2}(x^{5})\delta ^{\mu 2}-b_{3}^{-2}(x^{5})\delta ^{\mu 3}\right)
\smallskip
\end{gather}

Let us stress that metric (1) is supposed to hold at a {\em local} (and not
global) scale. We shall therefore refer to it as a ``{\em topical}''
deformed metric, because it is supposed to be valid not everywhere, but only
in a suitable (local) space-time region (characteristic of both the system
and the interaction considered).

The two basic postulates of DSR4 (which generalize those of standard SR) are
[1]:

\ \ 1- {\em Space-time properties: } Space-time is homogeneous, but space is
not necessarily isotropic; a reference frame in which space-time is endowed
with such properties is called a ''{\em topical'' reference frame} (TIRF).
Two TIRF's are in general moving uniformly with respect to each other (i.e.,
as in SR, they are connected by a ''inertiality'' relation, which defines an
equivalence class of $\infty ^{3}$ TIRF );

2- {\em Generalized Principle of Relativity }(or {\em Principle of Metric
Invariance}): All physical measurements within each TIRF must be carried out
via the {\em same} metric.

The metric (1) is just a possible realization of the above postulates. We
refer the reader to Ref.s [1]-[5] for the explicit expressions of the
phenomenological energy-dependent metrics for the four fundamental
interactions\footnote{%
Since the metric coefficients $b_{\mu }^{2}(x^{5})$ are {\em dimensionless},
they actually do depend on the ratio $\frac{x^{5}}{x_{0}^{5}}$, where
\[
x_{0}^{5}\equiv \ell _{0}E_{0}
\]
is a {\em fundamental length}, proportional (by the {\em %
dimensionally-transposing} constant $\ell _{0}$) to the {\em threshold energy%
} $E_{0}$, characteristic of the interaction considered (see Ref.s [1]-[5]).}%
.

\section{\ Maximal Killing group of N-d. generalized Minkowski spaces}

\subsection{Generalized Minkowski spaces}

We shall call {\em generalized Minkowski space }$\widetilde{M_{N}}(\left\{
x\right\} _{n.m.})$ a $N$-dimensional Riemann space with a global metric
structure determined by the (in general non-diagonal) metric tensor $g_{\mu
\nu }(\left\{ x\right\} _{n.m.})$ ($\mu ,\nu =1,2,...,N$), where $\left\{
x\right\} _{n.m.}$ denotes a set of $N_{n.m.}$ non-metrical coordinates
(i.e. different from the $N$ coordinates related to the dimensions of the
space considered). The metric interval in $\widetilde{M_{N}}(\left\{
x\right\} _{n.m.})$ therefore reads (ESC on throughout)
\begin{equation}
ds^{2}=g_{\mu \nu }(\left\{ x\right\} _{n.m.})dx^{\mu }dx^{\nu }.
\end{equation}
We shall assume the signature $(T,S)$ ($T$ timelike dimensions and $S=N-T$
spacelike dimensions). It follows that $\widetilde{M_{N}}(\left\{ x\right\}
_{n.m.})$ is {\em flat}, because {\em all} the components of the
Riemann-Christoffel tensor vanish.

Of course, an example is just provided by the 4-d. deformed Minkowski space $%
\widetilde{M_{4}}(x^{5})$.

\bigskip

\subsection{Killing equations in a generalized Minkowski space}

The Lie algebra of isometrical transformations of a $N$-dimensional
Riemannian space is determined by the solutions of the $N(N+1)/2$ Killing
equations
\begin{equation}
\xi _{\mu }(\underline{x})_{;\nu }+\xi _{\nu }(\underline{x})_{;\mu
}=0\smallskip \Longleftrightarrow \xi _{\lbrack \mu }(\underline{x})_{;\nu
]}=0,  \label{Killing}
\end{equation}
where the bracket $\left[ ...\right] $ means symmetrization with respect to
the enclosed indices, $;\mu $ denotes, as usual, covariant derivation with
respect to the coordinate $x^{\mu }$, and the contravariant Killing vector\ $%
\xi ^{\mu }(\underline{x})$ is the infinitesimal vector of the general
coordinate transformation
\begin{equation}
\left( x^{\prime }\right) ^{\mu }=\left( x^{\mu }\right) ^{\prime }=x^{\mu
}+\delta x^{\mu }(\underline{x}),
\end{equation}
namely:
\begin{equation}
\delta x^{\mu }(\underline{x})=\xi ^{\mu }(\underline{x}).  \label{Killing2}
\end{equation}

In the case of a $N$-d. generalized Minkowski space $\widetilde{M_{N}}%
(\left\{ x\right\} _{n.m.})$, being (as noted above) a special case of a
Riemann space with constant (zero) curvature, we can state that it is a
maximally symmetric space, i.e. admits a maximal Killing group with $%
N(N+1)/2 $ parameters. Moreover, covariant derivative reduces to ordinary
ones ($,\mu =\partial /\partial x^{\mu }$), whence the Killing equations (%
\ref{Killing}) reduce to
\begin{equation}
\xi _{\lbrack \mu }(\underline{x})_{,\nu ]}=0\Longleftrightarrow \xi _{\mu }(%
\underline{x})_{,\nu }+\xi _{\nu }(\underline{x})_{,\mu }=0\smallskip
\Longleftrightarrow \frac{\partial \xi _{\mu }(\underline{x})}{\partial
x^{\nu }}+\frac{\partial \xi _{\nu }(\underline{x})}{\partial x^{\mu }}=0.
\label{Killing-flat}
\end{equation}

By virtue of the use of the omomorphic exponential mapping, any finite
element $g$ of a Lie group $G_{L}$ of order $M$, acting on $\widetilde{M_{N}}%
(\left\{ x\right\} _{n.m.})$, can be written in the exponential form

\begin{equation}
\text{ }g=\exp \left( \sum_{i=1}^{M}\alpha _{i}T^{i}\right) ,
\end{equation}
where $\left\{ T^{i}\right\} _{i=1...M}$ is the generator basis of the Lie
algebra of $G\medskip _{L}$, and $\left\{ \alpha _{i}\right\} _{i=1...M}\in
R^{M}$ ($\left\{ \alpha _{i}\right\} =\left\{ \alpha _{i}\right\} (g)$).

Therefore, by a series development of the exponential:

\begin{equation}
g=\exp \left( \sum_{i=1}^{M}\alpha _{i}T^{i}\right) =\sum_{k=0}^{\infty }%
\frac{1}{k!}\left( \sum_{i=1}^{M}\alpha _{i}(g)T^{i}\right) ^{k},\text{ \ \ }
\end{equation}
and thus we get, for an infinitesimal element ($g\rightarrow \delta g$) ($%
\Leftrightarrow \left\{ \alpha _{i}(g)\right\} _{i=1...M}\in
R^{M}\rightarrow \left\{ \alpha _{i}(g)\right\} _{i=1...M}\in I_{\underline{0%
}}$ $\subset R^{M}$ ):
\begin{equation}
\delta g=1+\sum_{i=1}^{M}\alpha _{i}(g)T^{i}+O(\left\{ \alpha
_{i}^{2}(g)\right\} ).\text{ }
\end{equation}

Then, $\forall \underline{x}\in \widetilde{M_{N}}(\left\{ x\right\} _{n.m.})
$, one has, for the action of a finite and an infinitesimal element of $%
G_{L} $, respectively:

\begin{equation}
g\underline{x}=\left[ \exp \left( \sum_{i=1}^{M}\alpha _{i}(g)T^{i}\right) %
\right] \underline{x}=\left[ \sum_{k=0}^{\infty }\frac{1}{k!}\left(
\sum_{i=1}^{M}\alpha _{i}(g)T^{i}\right) ^{k}\right] \underline{x}=%
\underline{x^{\prime }}\in \widetilde{M_{N}}(\left\{ x\right\} _{n.m.})
\end{equation}

\begin{gather}
\left.
\begin{array}{l}
\left( \delta g\right) \underline{x}=\left[ 1+\sum_{i=1}^{M}\alpha
_{i}(g)T^{i}\right] \underline{x}=\underline{x}+\left( \sum_{i=1}^{M}\alpha
_{i}(g)T^{i}\right) \underline{x}=\underline{x^{\prime }}\in \widetilde{M_{N}%
}(\left\{ x\right\} _{n.m.}) \\
\\
\\
\delta g:\widetilde{M_{N}}(\left\{ x\right\} _{n.m.})\ni \underline{x}%
\rightarrow \underline{x^{\prime }}=\underline{x}+\underline{\delta x}_{(g)}(%
\underline{x})\in \widetilde{M_{N}}(\left\{ x\right\} _{n.m.})\text{\ \ \ \
\ \ \smallskip }
\end{array}
\right\} \medskip \Rightarrow  \nonumber \\
\nonumber \\
\Rightarrow \underline{\delta x}_{(g)}(\underline{x})=\left(
\sum_{i=1}^{M}\alpha _{i}(g)T^{i}\right) \underline{x}
\label{infinitesimal1}
\end{gather}

In index notation, Eq. (\ref{infinitesimal1}) can be written as

\begin{equation}
\delta x_{(g)}^{\mu }(\underline{x})=\left[ \left( \sum_{i=1}^{M}\alpha
_{i}(g)T^{i}\right) \underline{x}\right] ^{\mu },\mu =1,...,N;
\label{infinitesimal2}
\end{equation}
moreover, from (\ref{Killing2}) one gets

\begin{equation}
\xi _{(g)}^{\mu }(\underline{x})=\left[ \left( \sum_{i=1}^{M}\alpha
_{i}(g)T^{i}\right) \underline{x}\right] ^{\mu }.
\end{equation}

We can now define the mixed 2-rank N-tensor $\delta \omega _{~~~\nu }^{\mu
}(g,\left\{ x\right\} _{n.m.})$ of infinitesimal transformation (associated
to $\delta g\in G_{L}$) as (ESC on throughout):
\begin{equation}
\delta x_{(g)}^{\mu }(\underline{x},\left\{ x\right\} _{n.m.})=\left[ \left(
\sum_{i=1}^{M}\alpha _{i}(g)T^{i}(\left\{ x\right\} _{n.m.})\right)
\underline{x}\right] ^{\mu }\equiv \delta \omega _{~~~\nu }^{\mu }(g,\left\{
x\right\} _{n.m.})x^{\nu }.  \label{infinitesimal3}
\end{equation}
The number of independent components of tensor $\delta \omega _{~~~\nu
}^{\mu }(g,\left\{ x\right\} _{n.m.})$ is equal to the order $M$ of the Lie
group; in general, nothing can be said about its symmetry properties. From
Eq.s (\ref{infinitesimal2})-(\ref{infinitesimal3}) it follows that
\begin{equation}
\xi _{(g)}^{\mu }(\underline{x},\left\{ x\right\} _{n.m.})=\delta \omega
_{~~~\nu }^{\mu }(g,\left\{ x\right\} _{n.m.})x^{\nu },
\label{infinitesimal4}
\end{equation}
which shows that $\delta \omega _{~~~\nu }^{\mu }(g,\left\{ x\right\}
_{n.m.})$ is the tensor of the rotation parameters in $\widetilde{M_{N}}%
(\left\{ x\right\} _{n.m.})$\footnote{%
In the following, for simplicity of notation, we shall often omit the
explicit dependence of quantities on the non-metrical coordinates $\left\{
x\right\} _{n.m.}$.}.

Since we are looking for the Killing groups of $\widetilde{M_{N}}(\left\{
x\right\} _{n.m.})$ (not necessarily maximal), we have (from Eq.s (\ref
{infinitesimal1}), (\ref{infinitesimal2}) and (\ref{infinitesimal3})) :

\begin{gather}
\xi _{(g)\mu }(\underline{x})_{,\nu }+\xi _{(g)\nu }(\underline{x})_{,\mu
}=0\Leftrightarrow \delta x_{(g)\mu }(\underline{x})_{,\nu }+\delta
x_{(g)\nu }(\underline{x})_{,\mu }=0\Leftrightarrow  \nonumber \\
\nonumber \\
\Leftrightarrow \left[ \left( \sum_{i=1}^{M}\alpha _{i}(g)T^{i}\right)
\underline{x}\right] _{\mu ,\nu }+\left[ \left( \sum_{i=1}^{M}\alpha
_{i}(g)T^{i}\right) \underline{x}\right] _{\nu ,\mu }=0\Leftrightarrow
\nonumber \\
\medskip  \nonumber \\
\Leftrightarrow \left( \delta \omega _{\mu \rho }(g)x^{\rho }\right) _{,\nu
}+\left( \delta \omega _{\nu \rho }(g)x^{\rho }\right) _{,\mu
}=0\Leftrightarrow \medskip  \nonumber \\
\nonumber \\
\Leftrightarrow \left( \delta \omega _{\lbrack \mu \rho }(g)x^{\rho }\right)
_{,\nu ]}=0\smallskip .  \label{infinitesimal5}
\end{gather}
The last equation implies {\em antisymmetry} of $\delta \omega _{\mu \nu
}(g) $:

\begin{equation}
\delta \omega _{\mu \nu }(g)+\delta \omega _{\nu \mu }(g)=0\smallskip ,
\label{antisymm}
\end{equation}
which therefore has $N(N-1)/2$ independent components (such a number, as
stressed above, is also equal to to the order $M$ of $G_{L}$,i.e. $%
M=N(N-1)/2 $), i.e. the (rotation) transformation group related to the
tensor $\delta \omega _{\mu \nu }(g)$ is a $N(N-1)/2$-parameter Killing
group.

Since a $N$-d. generalized Minkowski space is maximally symmetric, we have
still to find another $N$-parameter Killing group of $\widetilde{M_{N}}%
(\left\{ x\right\} _{n.m.})$ (because $N+N(N-1)/2=N(N+1)/2$).

This is easily done by noting that the $N(N+1)/2$ Killing equations (\ref
{Killing-flat}) in such a space are trivially satisfied by constant
covariant $N$-vectors $\xi _{\mu }\neq \xi _{\mu }(\underline{x})$, to which
there corresponds the infinitesimal transformation

\begin{gather}
\delta g:x^{\mu }\rightarrow \left( x^{\prime }\right) ^{\mu }(\underline{x}%
,\left\{ x\right\} _{n.m.})=\left( x^{\mu }\right) ^{\prime }(\underline{x}%
,\left\{ x\right\} _{n.m.})=  \nonumber \\
\nonumber \\
=x^{\mu }+\delta x_{(g)}^{\mu }(\left\{ x\right\} _{n.m.})=x^{\mu }+\xi
_{(g)}^{\mu }(\left\{ x\right\} _{n.m.}),
\end{gather}
with $\delta x_{(g)}^{\mu }(\left\{ x\right\} _{n.m.})$ and $\xi _{(g)}^{\mu
}(\left\{ x\right\} _{n.m.})$ constant (with respect to $x^{\mu }$).

In conclusion, a {\em N}-d. generalized Minkowski space $\widetilde{M_{N}}%
(\left\{ x\right\} _{n.m.})$ admits a maximal Killing group which is the
(semidirect) product of the Lie group of {\em N}-dimensional space-time
rotations (or {\em N}-d. generalized, homogeneous Lorentz group $%
SO(T,S)_{GEN.}^{N(N-1)/2\text{{\footnotesize \ }}}$) with $N(N-1)/2$
parameters, and of the Lie group of {\em N}-dimensional space-time
translations $Tr.(T,S)_{GEN.}^{N}$ with $N$ parameters :

\begin{equation}
P(T,S)_{GEN.}^{N(N+1)/2\text{{\footnotesize \ }}}=SO(T,S)_{GEN.}^{N(N-1)/2}%
\otimes _{s}Tr.(T,S)_{GEN.}^{N\text{{\small \ }}}.
\end{equation}
We will refer to it as the {\em generalized Poincar\'{e}} (or {\em %
inhomogeneous Lorentz}) {\em group} $P(S,T)_{GEN.}^{N(N+1)/2}$.

\subsubsection{Solving the Killing equations in a 4-d. generalized Minkowski
space}

We want now to find the explicit solutions of the Killing equations in a
4-d. generalized Minkowski space $\widetilde{M_{4}}(\left\{ x\right\}
_{n.m.})$ \ ($S\leq 4,T=4-S$). A covariant Killing 4-vector $\xi _{\mu
}(x^{0},x^{1},x^{2},x^{3})$ must satisfy Eq. (\ref{Killing-flat}), namely
\begin{equation}
\left\{
\begin{array}{l}
I.\text{ \ \ }\dfrac{\partial \xi _{0}(\left\{ x\right\} _{m.})}{\partial
x^{0}}=0\smallskip \\
\\
II.\text{ \ \ }\dfrac{\partial \xi _{0}(\left\{ x\right\} _{m.})}{\partial
x^{1}}+\dfrac{\partial \xi _{1}(\left\{ x\right\} _{m.})}{\partial x^{0}}=0
\\
\smallskip \\
III.\text{ \ \ }\dfrac{\partial \xi _{0}(\left\{ x\right\} _{m.})}{\partial
x^{2}}+\dfrac{\partial \xi _{2}(\left\{ x\right\} _{m.})}{\partial x^{0}}%
=0\smallskip \\
\\
IV.\text{ \ \ }\dfrac{\partial \xi _{0}(\left\{ x\right\} _{m.})}{\partial
x^{3}}+\dfrac{\partial \xi _{3}(\left\{ x\right\} _{m.})}{\partial x^{0}}=0
\\
\smallskip \\
V.\text{ \ \ }\dfrac{\partial \xi _{1}(\left\{ x\right\} _{m.})}{\partial
x^{1}}=0\smallskip \\
\\
VI.\text{ \ \ }\dfrac{\partial \xi _{1}(\left\{ x\right\} _{m.})}{\partial
x^{2}}+\dfrac{\partial \xi _{2}(\left\{ x\right\} _{m.})}{\partial x^{1}}=0
\\
\smallskip \\
VII.\text{ \ \ }\dfrac{\partial \xi _{1}(\left\{ x\right\} _{m.})}{\partial
x^{3}}+\dfrac{\partial \xi _{3}(\left\{ x\right\} _{m.})}{\partial x^{1}}%
=0\smallskip \\
\\
VIII.\text{ \ \ }\dfrac{\partial \xi _{2}(\left\{ x\right\} _{m.})}{\partial
x^{2}}=0\smallskip \\
\\
IX.\text{ \ \ }\dfrac{\partial \xi _{2}(\left\{ x\right\} _{m.})}{\partial
x^{3}}+\dfrac{\partial \xi _{3}(\left\{ x\right\} _{m.})}{\partial x^{2}}%
=0\smallskip \\
\\
X.\text{ \ \ }\dfrac{\partial \xi _{3}(\left\{ x\right\} _{m.})}{\partial
x^{3}}=0\smallskip
\end{array}
\right.  \label{Killing-flat-4d}
\end{equation}

From Eq.s $I.$ ,$V.$,$VIII.$ and $X.$ one trivially gets:
\begin{equation}
\left\{
\begin{array}{c}
\xi _{0}=\xi _{0}(x^{1},x^{2},x^{3})\smallskip \\
\\
\xi _{1}=\xi _{1}(x^{0},x^{2},x^{3})\smallskip \\
\\
\xi _{2}=\xi _{2}(x^{0},x^{1},x^{3}) \\
\\
\xi _{3}=\xi _{3}(x^{0},x^{1},x^{2})\smallskip
\end{array}
\right. .
\end{equation}
In general, solving the system (\ref{Killing-flat-4d}) of 10 coupled partial
differential equations (PDEs) in 4 functional unknowns $\left( \xi _{0},\xi
_{1},\xi _{2},\xi _{3}\right) $ of 4 independent variables $(\left\{
x\right\} _{m.})=\left( x^{0},x^{1},x^{2},x^{3}\right) $ is cumbersome, but
straightforward. The final result is:

\begin{equation}
\left\{
\begin{array}{c}
\xi _{0}(\left\{ x\right\} _{m.})=-\zeta ^{1}x^{1}-\zeta ^{2}x^{2}-\zeta
^{3}x^{3}+T^{0}\medskip \\
\\
\xi _{1}(\left\{ x\right\} _{m.})=\zeta ^{1}x^{0}+\theta ^{2}x^{3}-\theta
^{3}x^{2}-T^{1}\medskip \\
\\
\xi _{2}(\left\{ x\right\} _{m.})=\zeta ^{2}x^{0}-\theta ^{1}x^{3}+\theta
^{3}x^{1}-T^{2}\medskip \\
\\
\xi _{3}(\left\{ x\right\} _{m.})=\zeta ^{3}x^{0}+\theta ^{1}x^{2}-\theta
^{2}x^{1}-T^{3}\medskip
\end{array}
\right.  \label{sol-flat}
\end{equation}
where $\zeta ^{i}$, $\theta ^{i}$ ($i=1,2,3$) and $T^{\mu }$ ($\mu =0,1,2,3$%
) are real coefficients.

Thus, we can draw the following conclusions:

{\bf 1}- In spite of the fact that no assumption was made on the functional
form of the Killing vector, we got a dependence at most linear
(inhomogeneous) on metric coordinates for all components of $\xi _{\mu
}(\left\{ x\right\} _{m.})$. Therefore, in order to determine the (maximal)
Killing group of a generalized Minkowski space \footnote{%
Indeed, although we discussed explicitly the 4-d. case, the extension to the
generic $N$-d. case is straightforward.}, one can, without loss of
generality, consider only groups whose transformation representation is
implemented by transformations at most linear in the coordinates.

{\bf 2}- In general, $\xi _{\mu }$ $\neq \xi _{\mu }(\left\{ x\right\}
_{n.m.})$, i.e. the covariant Killing vector does not depend on possible
non-metric variables. On the contrary, the {\em contravariant} Killing
4-vector {\em does indeed}, due to dependence of the fully contravariant
metric tensor on $\left\{ x\right\} _{n.m.}$ :
\begin{equation}
\xi ^{\mu }(\left\{ x\right\} _{m.},\left\{ x\right\} _{n.m.})=g^{\mu \nu
}(\left\{ x\right\} _{n.m.})\xi _{\nu }(\left\{ x\right\} _{m.}).
\label{sol-flat-controv}
\end{equation}
Such a result is consistent with the fact that $\delta \omega _{\mu \nu }(g)$%
, unlike $\delta \omega _{~~\nu }^{\mu }(g,\left\{ x\right\} _{n.m.})$, is
independent of $\left\{ x\right\} _{n.m.}$ (cfr. Eq.s (\ref{infinitesimal4})
and (\ref{infinitesimal5})).

{\bf 3} - Solution (\ref{sol-flat}) {\em does not depend on the metric tensor%
}. This implies that all 4-d. generalized Minkowski spaces admit the same
{\em covariant} Killing 4-vector. {\em It corresponds to the covariant
4-vector of infinitesimal transformation of the space-time
roto-translational group of }$\widetilde{M_{4}}(\left\{ x\right\} _{n.m.})$
. Therefore, by (\ref{sol-flat}) and (\ref{sol-flat-controv}), and assuming
the signature hyperbolic $(+,-,-,-)$ (i.e. $S=3$, $T=1$), in a basis of
''length-dimensional'' coordinates, we can state that:

{\bf 3.a}) $\underline{\zeta }=(\zeta ^{1},\zeta ^{2},\zeta ^{3})$ is the
3-vector of dimensionless parameters (''rapidity'') of generalized 3-d.
boost;

{\bf 3.b}) $\underline{\theta }=(\theta ^{1},\theta ^{2},\theta ^{3})$ is
the 3-vector of dimensionless parameters (angles) of generalized 3-d.
rotation;

{\bf 3.c}) $T_{\mu }=(T^{0},-T^{1},-T^{2},-T^{3})$ is the covariant 4-vector
of (''length-dimensional'') parameters of generalized 4-d. translation.

\section{Infinitesimal structure of generalized space-time rotation groups}

\subsection{Finite-dimensional representation of infinitesimal generators
and generalized Lorentz algebra}

As in the standard special-relatistic case, we can decompose the mixed $N$%
-tensor of infinitesimal transformation parameters $\delta \omega _{~~~\nu
}^{\mu }(g,\left\{ x\right\} _{n.m.})$ (see (\ref{infinitesimal3})) as
\footnote{%
The factor $\frac{1}{2}$\ is inserted only for further convenience.}:
\begin{equation}
\delta \omega _{~~\nu }^{\mu }(g,\left\{ x\right\} _{n.m.})=\frac{1}{2}%
\delta \omega _{\alpha \beta }(g)(I^{\alpha \beta })_{~~\nu }^{\mu }(\left\{
x\right\} _{n.m.})  \label{decomp0}
\end{equation}
i.e. as a linear combination of $N(N-1)/2$ matrices (indipendent of the
group element $g$) $\left\{ (I^{\alpha \beta })_{~~\nu }^{\mu }(\left\{
x\right\} _{n.m.})\right\} _{\alpha ,\beta =1...N}$ \footnote{%
The pair of indices $(\alpha $,$\beta )$ labels the different infinitesimal
group generators, whereas - in the ($N(<\infty )$-dimensional) matrix
representation of the generators we are considering - the controvariant
(covariant) index is a row (column) index. This latter remark holds true for
$\delta \omega _{~~\nu }^{\mu }(g,\left\{ x\right\} _{n.m.})$, too.} with
coefficients $\left\{ \delta \omega _{\alpha \beta }(g)\right\} _{\alpha
,\beta =1...N}$. Such matrices represent{\em \ the infinitesimal generators }%
of the space-time rotational component of the maximal Killing group of $%
\widetilde{M_{N}}(\left\{ x\right\} _{n.m.})$. Since in this case $\delta
\omega _{\mu \nu }(g)$ is antisymmetric (see (\ref{antisymm})), also the
basis matrices $\left\{ (I^{\alpha \beta })_{~~\nu }^{\mu }(\left\{
x\right\} _{n.m.})\right\} _{\alpha ,\beta =1...N}$ , are antisymmetric in
indices $\alpha $ and $\beta $ :
\begin{equation}
\left\{ (I^{\alpha \beta })_{~~\nu }^{\mu }(\left\{ x\right\}
_{n.m.})\right\} _{\alpha ,\beta =1...N}=-\left\{ (I^{\beta \alpha })_{~~\nu
}^{\mu }(\left\{ x\right\} _{n.m.})\right\} _{\alpha ,\beta =1...N}.
\end{equation}

For the fully covariant $\delta \omega _{\mu \nu }(g)$ the analogous
decomposition reads
\begin{gather}
\delta \omega _{\mu \nu ~}(g)=g_{\mu \rho }(\left\{ x\right\} _{n.m.})\delta
\omega _{~~\nu }^{\rho }(g,\left\{ x\right\} _{n.m.})=  \nonumber \\
\smallskip  \nonumber \\
=\frac{1}{2}\delta \omega _{\alpha \beta }(g)g_{\mu \rho }(\left\{ x\right\}
_{n.m.})(I^{\alpha \beta })_{~~\nu }^{\rho }(\left\{ x\right\} _{n.m.})=
\nonumber \\
\smallskip  \nonumber \\
=\frac{1}{2}\delta \omega _{\alpha \beta }(g)(I^{\alpha \beta })_{\mu \nu
}(\left\{ x\right\} _{n.m.})\smallskip .  \label{decomp1}
\end{gather}
But, since $\delta \omega _{\mu \nu }(g)$ is independent of $\left\{
x\right\} _{n.m.}$, the same holds for its components $(I^{\alpha \beta
})_{\mu \nu }$, and therefore Eq. (\ref{decomp1}) implies
\begin{equation}
(I^{\alpha \beta })_{\mu \nu }\neq (I^{\alpha \beta })_{\mu \nu }(\left\{
x\right\} _{n.m.}).
\end{equation}

In order to find the explicit form of the infinitesimal generators in the $N$%
-d. matrix representation, let us exploit the antisymmetry of $\delta \omega
_{\mu \nu ~}(g)$ :

\begin{gather}
\delta \omega _{\mu \nu ~}(g)=-\delta \omega _{\nu \mu ~}(g)\Leftrightarrow
\delta \omega _{\mu \nu ~}(g)=\smallskip  \nonumber \\
\nonumber \\
=\frac{1}{2}(\delta \omega _{\mu \nu ~}+\delta \omega _{\mu \nu ~})=\frac{1}{%
2}(\delta \omega _{\mu \nu ~}-\delta \omega _{\nu \mu ~})=\smallskip
\nonumber \\
\nonumber \\
=\frac{1}{2}g_{~~\mu }^{\alpha }g_{~~\nu }^{\beta }\delta \omega _{\alpha
\beta }-\frac{1}{2}g_{~~\mu }^{\beta }g_{~~\nu }^{\alpha }\delta \omega
_{\alpha \beta }=\frac{1}{2}\delta \omega _{\alpha \beta }(g_{~~\mu
}^{\alpha }g_{~~\nu }^{\beta }-g_{~~\mu }^{\beta }g_{~~\nu }^{\alpha
})\smallskip =  \nonumber \\
\nonumber \\
=\frac{1}{2}\delta \omega _{\alpha \beta }(g)(\delta _{~\ \mu }^{\alpha
}\delta _{~~\nu }^{\beta }-\delta _{~~\mu }^{\beta }\delta _{~~\nu }^{\alpha
}).  \label{decomp2}
\end{gather}
Comparing (\ref{decomp1}) and (\ref{decomp2}) yields \footnote{%
Eq. (\ref{decomp3}) clearly shows that the factors with a non-metric
dependence in $g_{\mu \rho }(\left\{ x\right\} _{n.m.})$ e $(I^{\alpha \beta
})_{~~\nu }^{\rho }(\left\{ x\right\} _{n.m.})$ annul each other. The same
{\em does not }happen when both $\mu $\ and $\nu $\ are controvariant:
\begin{gather*}
(I^{\alpha \beta })^{\mu \nu }:=g^{\mu \rho }(\left\{ x\right\}
_{n.m.})g^{\nu \sigma }(\left\{ x\right\} _{n.m.})(I^{\alpha \beta })_{\rho
\sigma }= \\
\\
=g^{\mu \rho }(\left\{ x\right\} _{n.m.})g^{\nu \sigma }(\left\{ x\right\}
_{n.m.})(\delta _{\rho }^{\alpha }\delta _{\sigma }^{\beta }-\delta _{\rho
}^{\beta }\delta _{\sigma }^{\alpha })= \\
\\
=g^{\mu \alpha }(\left\{ x\right\} _{n.m.})g^{\nu \beta }(\left\{ x\right\}
_{n.m.})-g^{\mu \beta }(\left\{ x\right\} _{n.m.})g^{\nu \alpha }(\left\{
x\right\} _{n.m.}).
\end{gather*}
\
\par
{}} :
\begin{equation}
g_{\mu \rho }(\left\{ x\right\} _{n.m.})(I^{\alpha \beta })_{~~\nu }^{\rho
}(\left\{ x\right\} _{n.m.})\equiv (I^{\alpha \beta })_{\mu \nu }=(\delta
_{~\ \mu }^{\alpha }\delta _{~~\nu }^{\beta }-\delta _{~~\mu }^{\beta
}\delta _{~~\nu }^{\alpha });  \label{decomp3}
\end{equation}
we get therefore the following explicit form for the mixed matrix
representation of the generators \footnote{%
We have analogously
\begin{gather*}
(I^{\alpha \beta })_{\mu }^{~~\nu }(\left\{ x\right\} _{n.m.})=g^{\nu \rho
}(\left\{ x\right\} _{n.m.})(I^{\alpha \beta })_{\mu \rho }=\smallskip \\
\\
=g^{\nu \rho }(\left\{ x\right\} _{n.m.})(\delta _{~\ \mu }^{\alpha }\delta
_{~~\rho }^{\beta }-\delta _{~~\mu }^{\beta }\delta _{~~\rho }^{\alpha
})=\smallskip \\
\\
=g^{\nu \beta }(\left\{ x\right\} _{n.m.})\delta _{~~\mu }^{\alpha }-g^{\nu
\alpha }(\left\{ x\right\} _{n.m.})\delta _{~~\mu }^{\beta }\smallskip .
\end{gather*}
}:
\begin{gather}
(I^{\alpha \beta })_{~~\nu }^{\mu }(\left\{ x\right\} _{n.m.})=g^{\mu \rho
}(\left\{ x\right\} _{n.m.})(I^{\alpha \beta })_{\rho \nu }=\smallskip
\nonumber \\
\nonumber \\
=g^{\mu \rho }(\left\{ x\right\} _{n.m.})(\delta _{~\ \rho }^{\alpha }\delta
_{~~\nu }^{\beta }-\delta _{~~\rho }^{\beta }\delta _{~~\nu }^{\alpha
})=g^{\mu \alpha }(\left\{ x\right\} _{n.m.})\delta _{~~\nu }^{\beta
}-g^{\mu \beta }(\left\{ x\right\} _{n.m.})\delta _{~~\nu }^{\alpha
}\smallskip .  \label{expr3}
\end{gather}

It is easy to see that the generators $\left\{ (I^{\alpha \beta })_{~~\nu
}^{\mu }(\left\{ x\right\} _{n.m.})\right\} _{\alpha ,\beta =1...N}$ satisfy
the following Lie algebra (with the usual commutatorial implementation of
Lie algebra product):
\begin{gather}
\lbrack I^{\alpha \beta }(\left\{ x\right\} _{n.m.}),I^{\rho \sigma
}(\left\{ x\right\} _{n.m.})]=\medskip  \nonumber \\
\nonumber \\
=g^{\alpha \sigma }(\left\{ x\right\} _{n.m.})I^{\beta \rho }(\left\{
x\right\} _{n.m.})+g^{\beta \rho }(\left\{ x\right\} _{n.m.})I^{\alpha
\sigma }(\left\{ x\right\} _{n.m.})+\medskip  \nonumber \\
\nonumber \\
-g^{\alpha \rho }(\left\{ x\right\} _{n.m.})I^{\beta \sigma }(\left\{
x\right\} _{n.m.})-g^{\beta \sigma }(\left\{ x\right\} _{n.m.})I^{\alpha
\rho }(\left\{ x\right\} _{n.m.})\smallskip .  \label{gen-Lor-1}
\end{gather}
Eq. (\ref{gen-Lor-1}) defines the {\em generalized Lorentz algebra},
associated to the generalized, homogeneous Lorentz group $%
SO(T,S)_{GEN.}^{N(N-1)/2}$ of the $N$-d. generalized Minkowski space $%
\widetilde{M_{N}}(\left\{ x\right\} _{n.m.})$.

\subsection{The case of a 4-dimensional generalized Minkowski space}

\subsubsection{ Self-representation of the infinitesimal generators}

Let us specialize the results of the previous Subsection to a 4-d.
generalized Minkowski space. Assuming therefore that Greek indices range in $%
\left\{ 0,1,2,3\right\} $, and that a (not necessarily hyperbolic) signature
$(S\leqslant 4,T=4-S)$ holds, we can write explicitly the generator $%
I^{\alpha \beta }(\left\{ x\right\} _{n.m.})$ of $SO(S,T=4-S)_{GEN}$ as the
antisymmetric matrix:
\begin{gather}
I^{\alpha \beta }(\left\{ x\right\} _{n.m.})=  \nonumber \\
\nonumber \\
=\left(
\begin{array}{cccc}
0 & I^{01}(\left\{ x\right\} _{n.m.}) & I^{02}(\left\{ x\right\} _{n.m.}) &
I^{03}(\left\{ x\right\} _{n.m.}) \\
-I^{01}(\left\{ x\right\} _{n.m.}) & 0 & I^{12}(\left\{ x\right\} _{n.m.}) &
I^{13}(\left\{ x\right\} _{n.m.}) \\
-I^{02}(\left\{ x\right\} _{n.m.}) & -I^{12}(\left\{ x\right\} _{n.m.}) & 0
& I^{23}(\left\{ x\right\} _{n.m.}) \\
-I^{03}(\left\{ x\right\} _{n.m.}) & -I^{13}(\left\{ x\right\} _{n.m.}) &
-I^{23}(\left\{ x\right\} _{n.m.}) & 0
\end{array}
\right) .
\end{gather}
Like any rank-2, antisymmetric 4-tensor, $I^{\alpha \beta }(\left\{
x\right\} _{n.m.})$ can be expressed in terms of an axial and a polar
3-vector. By introducing the following infinitesimal generators $%
(i,j,k=1,2,3 $, ESC on throughout)
\begin{gather}
S^{i}(\left\{ x\right\} _{n.m.})\equiv \frac{1}{2}\epsilon
_{~jk}^{i}I^{jk}(\left\{ x\right\} _{n.m.})\medskip  \label{S} \\
\nonumber \\
K^{i}(\left\{ x\right\} _{n.m.})\equiv I^{0i}(\left\{ x\right\} _{n.m.})
\label{K}
\end{gather}
(where $\epsilon _{ijk}$ is the rank-3, fully antisymmetric Levi-Civita
3-tensor, with the convention $\epsilon _{123}\equiv 1$), corresponding to
the components of the axial operatorial 3-vector
\begin{equation}
\underline{S}(\left\{ x\right\} _{n.m.})\equiv \smallskip (I^{23}(\left\{
x\right\} _{n.m.}),I^{31}(\left\{ x\right\} _{n.m.}),I^{12}(\left\{
x\right\} _{n.m.}))
\end{equation}
and of the polar operatorial one
\begin{equation}
\underline{K}(\left\{ x\right\} _{n.m.})\equiv (I^{01}(\left\{ x\right\}
_{n.m.}),I^{02}(\left\{ x\right\} _{n.m.}),I^{03}(\left\{ x\right\}
_{n.m.})),\text{ }
\end{equation}
$I^{\alpha \beta }(\left\{ x\right\} _{n.m.})$ can thus be rewritten as:
\begin{gather}
I^{\alpha \beta }(\left\{ x\right\} _{n.m.})=  \nonumber \\
\nonumber \\
=\left(
\begin{array}{cccc}
0 & K^{1}(\left\{ x\right\} _{n.m.}) & K^{2}(\left\{ x\right\} _{n.m.}) &
K^{3}(\left\{ x\right\} _{n.m.}) \\
-K^{1}(\left\{ x\right\} _{n.m.}) & 0 & S^{3}(\left\{ x\right\} _{n.m.}) &
-S^{2}(\left\{ x\right\} _{n.m.}) \\
-K^{2}(\left\{ x\right\} _{n.m.}) & -S^{3}(\left\{ x\right\} _{n.m.}) & 0 &
S^{1}(\left\{ x\right\} _{n.m.}) \\
-K^{3}(\left\{ x\right\} _{n.m.}) & S^{2}(\left\{ x\right\} _{n.m.}) &
-S^{1}(\left\{ x\right\} _{n.m.}) & 0
\end{array}
\right) .
\end{gather}

The set of generators $\underline{S}(\left\{ x\right\} _{n.m.})$ , $%
\underline{K}(\left\{ x\right\} _{n.m.})$ constitute the {\em %
self-representation basis }for $SO(S,T=4-S)_{GEN.}$. Unlike the case of
standard SR - where $\underline{S}$ , $\underline{K}$ do represent the
rotation and boost generators, respectively -, one cannot give them a
precise physical meaning, because this latter depends on both the number $S$
of spacelike dimensions and the assignement of dimensional labelling (for
full generality, here we left it unspecified, even if it is clear that the
most directly physically meaningful case is the hyperbolically signed-one: $%
S=3$).

\subsubsection{Decomposition of the parametric 4-tensor $\protect\delta
\protect\omega _{\protect\mu \protect\nu }(g)$}

We can now exploit the self-representation form of the infinitesimal
generators of $SO(S,T=4-S)_{GEN.}$ ($S\leq 4$) to decompose the
infinitesimal parametric 4-tensor $\delta \omega _{\mu \nu }(g)$.

Eq. (\ref{infinitesimal3}),\ on account of (\ref{decomp0}), can be written
as
\begin{equation}
\delta x_{(g)}^{\mu }(\left\{ x\right\} _{m.},\left\{ x\right\}
_{n.m.})=\delta \omega _{~~~\nu }^{\mu }(g,\left\{ x\right\} _{n.m.})x^{\nu
}=\frac{1}{2}\delta \omega _{\alpha \beta }(g)(I^{\alpha \beta })_{~~\nu
}^{\mu }(\left\{ x\right\} _{n.m.})x^{\nu },
\end{equation}
which is valid in the general case of $SO(S,T=N-S)_{GEN.}$.

In the case $N=4$ , we have (by (\ref{K})):
\begin{gather}
\delta x_{(g)}^{\mu }(\left\{ x\right\} _{m.},\left\{ x\right\}
_{n.m.})=\delta \omega _{~~~\nu }^{\mu }(g,\left\{ x\right\} _{n.m.})x^{\nu
}=\frac{1}{2}\delta \omega _{\alpha \beta }(g)(I^{\alpha \beta })_{~~\nu
}^{\mu }(\left\{ x\right\} _{n.m.})x^{\nu }=  \nonumber \\
\medskip   \nonumber \\
=\frac{1}{2}\delta \omega _{ij}(g)(I^{ij})_{~~\nu }^{\mu }(\left\{ x\right\}
_{n.m.})x^{\nu }+\delta \omega _{0i}(g)(I^{0i})_{~~\nu }^{\mu }(\left\{
x\right\} _{n.m.})x^{\nu }=\medskip   \nonumber \\
\nonumber \\
=\frac{1}{2}\delta \omega _{ij}(g)(I^{ij})_{~~\nu }^{\mu }(\left\{ x\right\}
_{n.m.})x^{\nu }+\delta \omega _{0i}(g)(K^{i})_{~~\nu }^{\mu }(\left\{
x\right\} _{n.m.})x^{\nu }\smallskip .  \label{expr1}
\end{gather}
Moreover, from (\ref{S}) we get:
\begin{equation}
S^{i}(\left\{ x\right\} _{n.m.})\equiv \frac{1}{2}\epsilon
_{~jk}^{i}I^{jk}(\left\{ x\right\} _{n.m.})\Leftrightarrow I^{jk}(\left\{
x\right\} _{n.m.})=\epsilon _{~~l}^{jk}S^{l}(\left\{ x\right\} _{n.m.});
\end{equation}
replacing such result in (\ref{expr1}) one thus has:
\begin{gather}
\delta x_{(g)}^{\mu }(\left\{ x\right\} _{m.},\left\{ x\right\} _{n.m.})=%
\frac{1}{2}\delta \omega _{ij}(g)(I^{ij})_{~~\nu }^{\mu }(\left\{ x\right\}
_{n.m.})x^{\nu }+\delta \omega _{0i}(g)(K^{i})_{~~\nu }^{\mu }(\left\{
x\right\} _{n.m.})x^{\nu }=\medskip   \nonumber \\
\nonumber \\
=\frac{1}{2}\delta \omega _{ij}(g)(\epsilon _{~~l}^{ij}S^{l}(\left\{
x\right\} _{n.m.}))_{~~\nu }^{\mu }x^{\nu }+\delta \omega
_{0i}(g)(K^{i})_{~~\nu }^{\mu }(\left\{ x\right\} _{n.m.})x^{\nu }=\medskip
\nonumber \\
\nonumber \\
=\frac{1}{2}\epsilon _{~~l}^{ij}\delta \omega _{ij}(g)(S^{l})_{~~\nu }^{\mu
}(\left\{ x\right\} _{n.m.})x^{\nu }+\delta \omega _{0i}(g)(K^{i})_{~~\nu
}^{\mu }(\left\{ x\right\} _{n.m.})x^{\nu }\smallskip .  \label{expr2}
\end{gather}
We can now define an axial and a polar parametric 3-vector in the following
way:
\begin{gather}
\theta _{i}(g)\equiv -\frac{1}{2}\epsilon _{i}^{~jk}\delta \omega
_{jk}(g)\medskip  \\
\nonumber \\
\zeta _{i}(g)\equiv -\delta \omega _{0i}(g),
\end{gather}
namely
\begin{equation}
\underline{\theta }(g)\equiv \text{{\footnotesize \ }}(-\delta \omega
_{23}(g),-\delta \omega _{31}(g),-\delta \omega _{12}(g))\text{%
{\footnotesize \ }}  \label{Theta}
\end{equation}
\begin{equation}
\underline{\zeta }(g)\equiv (-\delta \omega _{01}(g),-\delta \omega
_{02}(g),-\delta \omega _{03}(g)).\text{ }  \label{Zeta}
\end{equation}
Therefore, $\delta \omega _{\alpha \beta }(g)$ can be written in matrix form
as:
\begin{equation}
\begin{array}{cc}
\delta \omega _{\alpha \beta }(g)= & \left(
\begin{array}{cccc}
0 & -\zeta ^{1}(g) & -\zeta ^{2}(g) & -\zeta ^{3}(g) \\
\zeta ^{1}(g) & 0 & -\theta ^{3}(g) & \theta ^{2}(g) \\
\zeta ^{2}(g) & \theta ^{3}(g) & 0 & -\theta ^{1}(g) \\
\zeta ^{3}(g) & -\theta ^{2}(g) & \theta ^{1}(g) & 0
\end{array}
\right) .
\end{array}
\end{equation}

Having left the number $S$ of spacelike dimensions and the dimensional
labelling unspecified, we cannot attribute a physical meaning to the
parametric 3-vectors (\ref{Theta}),(\ref{Zeta}) (unlike the case of standard
SR, where $\underline{\theta }(g)$ and $\underline{\zeta }(g)$ are the space
rotation and boost \ parameters, respectively).

Eq. (\ref{expr2}) can therefore be rewritten in terms of the 3-d. Euclidean
scalar product $\cdot $ as:
\begin{gather}
\delta x_{(g)}^{\mu }(\left\{ x\right\} _{m.},\left\{ x\right\}
_{n.m.})=-\theta _{l}(g)(S^{l})_{~~\nu }^{\mu }(\left\{ x\right\}
_{n.m.})x^{\nu }-\zeta _{i}(g)(K^{i})_{~~\nu }^{\mu }\left\{ x\right\}
_{n.m.})x^{\nu }=  \nonumber \\
\medskip  \nonumber \\
=\left[ -\underline{\theta (g)}\cdot \underline{S}(\left\{ x\right\}
_{n.m.})-\underline{\zeta (g)}\cdot \underline{K}(\left\{ x\right\} _{n.m.})%
\right] _{~~\nu }^{\mu }x^{\nu }\smallskip
\end{gather}

\bigskip

\section{Space-time rotation component of the Killing group in a 4-d.
deformed Minkowski space}

\subsection{Deformed homogeneous Lorentz group $SO(3,1)_{DEF.}$ and
self-representation basis of infinitesimal generators}

We want now to specialize the results obtained to the case of DSR4, i.e.
considering a 4-d. deformed Minkowski space $\widetilde{M_{4}}(x^{5})$.

Let us recall that the $N$-dimensional representation of the infinitesimal
generators of the Killing group in a $N$-d. generalized Minkowski space is
determined (by means of eq. (\ref{expr3})) by the mere knowledge of its
metric tensor. In the DSR4 case we have therefore:
\begin{gather}
(I^{\alpha \beta })_{~~\nu ,DSR4}^{\mu }(x^{5})=g_{DSR4}^{\mu \rho
}(x^{5})(I^{\alpha \beta })_{\rho \nu ,DSR4}=\medskip  \nonumber \\
\nonumber \\
=g_{DSR4}^{\mu \rho }(x^{5})(\delta _{~\ \rho }^{\alpha }\delta _{~~\nu
}^{\beta }-\delta _{~~\rho }^{\beta }\delta _{~~\nu }^{\alpha
})=g_{DSR4}^{\mu \alpha }(x^{5})\delta _{~~\nu }^{\beta }-g_{DSR4}^{\mu
\beta }(x^{5})\delta _{~~\nu }^{\alpha }=  \nonumber \\
\nonumber \\
\stackrel{\text{{\footnotesize ESC off}}}{=}\delta ^{\mu \alpha }\left(
b_{0}^{-2}(x^{5})\delta ^{\mu 0}-b_{1}^{-2}(x^{5})\delta ^{\mu
1}-b_{2}^{-2}(x^{5})\delta ^{\mu 2}-b_{3}^{-2}(x^{5})\delta ^{\mu 3}\right)
\delta _{~~\nu }^{\beta }+\medskip  \nonumber \\
\nonumber \\
-\delta ^{\mu \beta }\left( b_{0}^{-2}(x^{5})\delta ^{\mu
0}-b_{1}^{-2}(x^{5})\delta ^{\mu 1}-b_{2}^{-2}(x^{5})\delta ^{\mu
2}-b_{3}^{-2}(x^{5})\delta ^{\mu 3}\right) \delta _{~~\nu }^{\alpha
}\smallskip .
\end{gather}
From such result, we thence get the following $4\times 4$ matrix
representation of the infinitesimal generator of the deformed homogeneous
Lorentz group $SO(3,1)_{DEF.}^{6}$ (space-time rotation component of the
deformed Poincar\'{e} group $P(3,1)_{DEF.}^{10}$):
\begin{equation}
\begin{array}{cc}
I_{DSR4}^{10}(x^{5})= & \left(
\begin{array}{cccc}
0 & -b_{0}^{-2}(x^{5}) & 0 & 0 \\
-b_{1}^{-2}(x^{5}) & 0 & 0 & 0 \\
0 & 0 & 0 & 0 \\
0 & 0 & 0 & 0
\end{array}
\right) ,\label{I10}
\end{array}
\end{equation}
\begin{equation}
\begin{array}{cc}
I_{DSR4}^{20}(x^{5})= & \left(
\begin{array}{cccc}
0 & 0 & -b_{0}^{-2}(x^{5}) & 0 \\
0 & 0 & 0 & 0 \\
-b_{2}^{-2}(x^{5}) & 0 & 0 & 0 \\
0 & 0 & 0 & 0
\end{array}
\right) ,
\end{array}
\end{equation}
\begin{equation}
\begin{array}{cc}
I_{DSR4}^{30}(x^{5})= & \left(
\begin{array}{cccc}
0 & 0 & 0 & -b_{0}^{-2}(x^{5}) \\
0 & 0 & 0 & 0 \\
0 & 0 & 0 & 0 \\
-b_{3}^{-2}(x^{5}) & 0 & 0 & 0
\end{array}
\right) ,
\end{array}
\end{equation}
\begin{equation}
\begin{array}{cc}
I_{DSR4}^{12}(x^{5})= & \left(
\begin{array}{cccc}
0 & 0 & 0 & 0 \\
0 & 0 & -b_{1}^{-2}(x^{5}) & 0 \\
0 & b_{2}^{-2}(x^{5}) & 0 & 0 \\
0 & 0 & 0 & 0
\end{array}
\right) ,\label{I12}
\end{array}
\end{equation}
\begin{equation}
\begin{array}{cc}
I_{DSR4}^{23}(x^{5})= & \left(
\begin{array}{cccc}
0 & 0 & 0 & 0 \\
0 & 0 & 0 & 0 \\
0 & 0 & 0 & -b_{2}^{-2}(x^{5}) \\
0 & 0 & b_{3}^{-2}(x^{5}) & 0
\end{array}
\right) ,
\end{array}
\end{equation}
\begin{equation}
\begin{array}{cc}
I_{DSR4}^{31}(x^{5})= & \left(
\begin{array}{cccc}
0 & 0 & 0 & 0 \\
0 & 0 & 0 & b_{1}^{-2}(x^{5}) \\
0 & 0 & 0 & 0 \\
0 & -b_{3}^{-2}(x^{5}) & 0 & 0
\end{array}
\right) .\label{I31}
\end{array}
\end{equation}
A comparison of Eqs. (\ref{I10})-(\ref{I31}) with the 4-d. matrix
representation of the infinitesimal generators of the standard homogeneous
Lorentz group $SO(3,1)$ shows that {\em the deformation of the metric
structure implies the loss of symmetry of the boost generators and of
antisymmetry of space-rotation generators}.

The antisymmetry of the generators in the labelling indices $(\alpha ,\beta
) $ still holds:
\begin{equation}
\left\{ (I^{\alpha \beta })_{~~\nu ,DSR4}^{\mu }(x^{5})\right\} _{\alpha
,\beta =0,1,2,3}=-\left\{ (I^{\beta \alpha })_{~~\nu ,DSR4}^{\mu
}(x^{5})\right\} _{\alpha ,\beta =0,1,2,3},
\end{equation}
or, equivalently:
\begin{equation}
I_{DSR4}^{\alpha \beta }(x^{5})=-I_{DSR4}^{\beta \alpha }(x^{5}),\text{ \ \ }%
\alpha ,\beta =0,1,2,3.
\end{equation}
Therefore, there are only 6 independent generators. In matrix form we get:
\begin{equation}
\begin{array}{cc}
I_{DSR4}^{\alpha \beta }(x^{5})= & \left(
\begin{array}{cccc}
0 & I_{DSR4}^{01}(x^{5}) & I_{DSR4}^{02}(x^{5}) & I_{DSR4}^{03}(x^{5}) \\
-I_{DSR4}^{01}(x^{5}) & 0 & I_{DSR4}^{12}(x^{5}) & I_{DSR4}^{13}(x^{5}) \\
-I_{DSR4}^{02}(x^{5}) & -I_{DSR4}^{12}(x^{5}) & 0 & I_{DSR4}^{23}(x^{5}) \\
-I_{DSR4}^{03}(x^{5}) & -I_{DSR4}^{13}(x^{5}) & -I_{DSR4}^{23}(x^{5}) & 0
\end{array}
\right) .
\end{array}
\end{equation}

We can now pass to the self-representation basis of the generators of $%
SO(3,1)_{DEF.}$ by introducing the following axial and polar 3-vectors by
means of the Levi-Civita tensor:
\begin{gather}
S_{DSR4}^{i}(x^{5})\equiv \frac{1}{2}\epsilon
_{~jk}^{i}I_{DSR4}^{jk}(x^{5})\medskip ,  \label{S-DSR4} \\
\nonumber \\
K_{DSR4}^{i}(x^{5})\equiv I_{DSR4}^{0i}(x^{5}),  \label{K-DSR4}
\end{gather}
or
\begin{equation}
\underline{S_{DSR4}}(x^{5})\equiv
(I_{DSR4}^{23}(x^{5}),I_{DSR4}^{31}(x^{5}),I_{DSR4}^{12}(x^{5})),\text{%
{\footnotesize \ }}
\end{equation}
\begin{equation}
\underline{K_{DSR4}}(x^{5})\equiv
(I_{DSR4}^{01}(x^{5}),I_{DSR4}^{02}(x^{5}),I_{DSR4}^{03}(x^{5})).
\end{equation}
Then, $I_{DSR4}^{\alpha \beta }(x^{5})$ can be written as:
\begin{gather}
\begin{array}{cc}
I_{DSR4}^{\alpha \beta }(x^{5})= & \left(
\begin{array}{cccc}
0 & K_{DSR4}^{1}(x^{5}) & K_{DSR4}^{2}(x^{5}) & K_{DSR4}^{3}(x^{5}) \\
-K_{DSR4}^{1}(x^{5}) & 0 & S_{DSR4}^{3}(x^{5}) & -S_{DSR4}^{2}(x^{5}) \\
-K_{DSR4}^{2}(x^{5}) & -S_{DSR4}^{3}(x^{5}) & 0 & S_{DSR4}^{1}(x^{5}) \\
-K_{DSR4}^{3}(x^{5}) & S_{DSR4}^{2}(x^{5}) & -S_{DSR4}^{1}(x^{5}) & 0
\end{array}
\right) .
\end{array}
\nonumber \\
\end{gather}

Due to the hyperbolicity of the assumed metric signature, in the DSR4, like
in the SR case, we can identify\ (apart from a sign) $S_{DSR4}^{i}(x^{5})$
with the infinitesimal generator of the deformed 3-d. space rotation around $%
\widehat{x^{i}}$ , and $K_{DSR4}^{i}(x^{5})$ with the infinitesimal
generator of the deformed Lorentz boost with motion direction along $%
\widehat{x^{i}}$.

\subsection{Decomposition of the parametric 4-tensor $\protect\delta \protect%
\omega _{\protect\mu \protect\nu }(g)$ in DSR4 \ \ \ }

We can now specialize Eq. (\ref{expr2}) (expressing the infinitesimal
variation of the contravariant 4-vector $x^{\mu }$ in the
self-representation) to the DSR4 case by using Eq.s (\ref{S-DSR4}) and (\ref
{K-DSR4}) (ESC on throughout):
\begin{gather}
\delta x_{(g),DSR4}^{\mu }(\left\{ x\right\} _{m.},x^{5})=\delta \omega
_{~~~\nu ,DSR4}^{\mu }(g,x^{5})x^{\nu }=\medskip  \nonumber \\
\nonumber \\
=\frac{1}{2}\delta \omega _{\alpha \beta ,(DSR4)}(g)(I^{\alpha \beta
})_{~~\nu ,DSR4}^{\mu }(x^{5})x^{\nu }=\medskip  \nonumber \\
\nonumber \\
=\frac{1}{2}\epsilon _{~~l}^{ij}\delta \omega _{ij,(DSR4)}(g)(S^{l})_{~~\nu
,DSR4}^{\mu }(x^{5})x^{\nu }+\delta \omega _{0i,(DSR4)}(g)(K^{i})_{~~\nu
,DSR4}^{\mu }(x^{5})x^{\nu }\smallskip .  \label{expr-DSR4}
\end{gather}
Therefore, the parametric 4-tensor $\delta \omega _{\alpha \beta }(g)$ can
be written as
\begin{equation}
\begin{array}{cc}
\delta \omega _{\alpha \beta }(g)= & \left(
\begin{array}{cccc}
0 & -\zeta ^{1}(g) & -\zeta ^{2}(g) & -\zeta ^{3}(g) \\
\zeta ^{1}(g) & 0 & -\theta ^{3}(g) & \theta ^{2}(g) \\
\zeta ^{2}(g) & \theta ^{3}(g) & 0 & -\theta ^{1}(g) \\
\zeta ^{3}(g) & -\theta ^{2}(g) & \theta ^{1}(g) & 0
\end{array}
\right) ,
\end{array}
\end{equation}
where the parameter axial 3-vector $\underline{\theta }(g)$ and the
parameter polar 3-vector $\underline{\zeta }(g)$ are respectively defined
by:
\begin{gather}
\underline{\theta }(g)=(\theta _{i}(g))\equiv \left( -\frac{1}{2}\epsilon
_{i}^{~jk}\delta \omega _{jk}(g)\medskip \right) =(-\delta \omega
_{23}(g),-\delta \omega _{31}(g),-\delta \omega _{12}(g)), \\
\nonumber \\
\underline{\zeta }(g)=\zeta _{i}(g)\equiv (-\delta \omega _{0i}(g))=(-\delta
\omega _{01}(g),-\delta \omega _{02}(g),-\delta \omega _{03}(g)).
\end{gather}
and correspond to a true deformed rotation and to a deformed boost,
respectively.

\subsection{The infinitesimal transformations of the 4-d., deformed
homogeneous Lorentz group $SO(3,1)_{DEF.}$.}

We can now use the results of the previous two Subsections to write Eq. (\ref
{expr-DSR4}) as:
\begin{gather}
\delta x_{(g),DSR4}^{\mu }(\left\{ x\right\} _{m.},x^{5})=-\theta
_{l}(g)(S^{l})_{~~\nu ,DSR4}^{\mu }(x^{5})x^{\nu }-\zeta
_{i}(g)(K^{i})_{~~\nu ,DSR4}^{\mu }(x^{5})x^{\nu }=  \nonumber \\
\medskip  \nonumber \\
=\left( -\underline{\theta }(g)\cdot \underline{S_{DSR4}}(x^{5})-\underline{%
\zeta }(g)\cdot \underline{K_{DSR4}}(x^{5})\right) _{~~\nu }^{\mu }x^{\nu
}\smallskip .
\end{gather}

Therefore, the infinitesimal space-time rotation transformation in the
deformed Minkowski space $\widetilde{M_{4}}(x^{5})$ corresponding to the
element $g$ of $SO(3,1)_{DEF.}$, can be expressed as:
\begin{gather}
\delta g:x^{\mu }\rightarrow \left( x^{\prime }\right) _{(g),DSR4}^{\mu
}(\left\{ x\right\} _{m.},x^{5})=  \nonumber \\
\nonumber \\
=\left( x^{\mu }\right) _{(g),DSR4}^{\prime }(\left\{ x\right\}
_{m.},x^{5})=x^{\mu }+\delta x_{(g),DSR4}^{\mu }(\left\{ x\right\}
_{m.},x^{5})=  \nonumber \\
\medskip   \nonumber \\
=(1-\theta _{1}(g)S_{DSR4}^{1}(x^{5})-\theta
_{2}(g)S_{DSR4}^{2}(x^{5})-\theta _{3}(g)S_{DSR4}^{3}(x^{5})+  \nonumber \\
\medskip   \nonumber \\
-\zeta _{1}(g)K_{DSR4}^{1}(x^{5})-\zeta _{2}(g)K_{DSR4}^{2}(x^{5})-\zeta
_{3}(g)K_{DSR4}^{3}(x^{5}))_{~~\nu }^{\mu }x^{\nu }\smallskip ,
\end{gather}
where $1$ is the identity of $SO(3,1)_{DEF.}$.

Then, on account of the physical meaning of the 3-d. parameter and generator
vectors (respectively $\underline{\theta }(g)$ , $\underline{\zeta }(g)$ ,
and $\underline{S_{DSR4}}(x^{5})$ , $\underline{K_{DSR4}}(x^{5})$), we can
get, from the matrix representation of the $SO(3,1)_{DEF.}$ generators, the
explicit expressions of all the different kinds of infinitesimal
transformations of the deformed Lorentz group, namely:

{\bf 1} -{\em \ 3-d. deformed space (true) rotations} (parameters $%
\underline{\theta }(g)$ and generators $\underline{S_{DSR4}}(x^{5})$), which
constitute the group $SO(3)_{DEF.}$ of rotations in a deformed 3-d. space,
non-Abelian, non-invariant proper\ subgroup of $SO(3,1)_{DEF.}$:

{\bf 1.1} - (Clockwise) infinitesimal rotation by an angle $\theta _{1}(g)$
around $\widehat{x^{1}}$:
\begin{gather}
\left( x^{\prime }\right) _{(g),DSR4}^{\mu }(\left\{ x\right\}
_{m.},x^{5})=\left( x^{\mu }\right) _{(g),DSR4}^{\prime }(\left\{ x\right\}
_{m.},x^{5})=(1-\theta _{1}(g)S_{DSR4}^{1})_{~~\nu }^{\mu }(x^{5})x^{\nu
}\Leftrightarrow   \nonumber \\
\nonumber \\
\nonumber \\
\Leftrightarrow \left(
\begin{array}{c}
x_{(g),DSR4}^{0\prime }(x^{5}) \\
x_{(g),DSR4}^{1\prime }(x^{5}) \\
x_{(g),DSR4}^{2\prime }(x^{5}) \\
x_{(g),DSR4}^{3\prime }(x^{5})
\end{array}
\right) =\left(
\begin{array}{cccc}
1 & 0 & 0 & 0 \\
0 & 1 & 0 & 0 \\
0 & 0 & 1 & \theta _{1}(g)b_{2}^{-2}(x^{5}) \\
0 & 0 & -\theta _{1}(g)b_{3}^{-2}(x^{5}) & 1
\end{array}
\right) \left(
\begin{array}{c}
x^{0} \\
x^{1} \\
x^{2} \\
x^{3}
\end{array}
\right) =  \nonumber \\
\medskip   \nonumber \\
=\left(
\begin{array}{c}
x^{0} \\
x^{1} \\
x^{2}+\theta _{1}(g)b_{2}^{-2}(x^{5})x^{3} \\
-\theta _{1}(g)b_{3}^{-2}(x^{5})x^{2}+x^{3}
\end{array}
\right) \smallskip ;
\end{gather}

{\bf 1.2} - (Clockwise) infinitesimal rotation by an angle $\theta _{2}(g)$
around $\widehat{x^{2}}$ :
\begin{gather}
\left( x^{\prime }\right) _{(g),DSR4}^{\mu }(\left\{ x\right\}
_{m.},x^{5})=\left( x^{\mu }\right) _{(g),DSR4}^{\prime }(\left\{ x\right\}
_{m.},x^{5})=(1-\theta _{2}(g)S_{DSR4}^{2})_{~~\nu }^{\mu }(x^{5})x^{\nu
}\Leftrightarrow   \nonumber \\
\nonumber \\
\nonumber \\
\Leftrightarrow \left(
\begin{array}{c}
x_{(g),DSR4}^{0\prime }(x^{5}) \\
x_{(g),DSR4}^{1\prime }(x^{5}) \\
x_{(g),DSR4}^{2\prime }(x^{5}) \\
x_{(g),DSR4}^{3\prime }(x^{5})
\end{array}
\right) =\left(
\begin{array}{cccc}
1 & 0 & 0 & 0 \\
0 & 1 & 0 & -\theta _{2}(g)b_{1}^{-2}(x^{5}) \\
0 & 0 & 1 & 0 \\
0 & \theta _{2}(g)b_{3}^{-2}(x^{5}) & 0 & 1
\end{array}
\right) \left(
\begin{array}{c}
x^{0} \\
x^{1} \\
x^{2} \\
x^{3}
\end{array}
\right) =  \nonumber \\
\medskip   \nonumber \\
=\left(
\begin{array}{c}
x^{0} \\
x^{1}-\theta _{2}(g)b_{1}^{-2}(x^{5})x^{3} \\
x^{2} \\
\theta _{2}(g)b_{3}^{-2}(x^{5})x^{1}+x^{3}
\end{array}
\right) \smallskip ;
\end{gather}

{\bf 1.3} - (Clockwise) infinitesimal rotation by an angle $\theta _{3}(g)$
around $\widehat{x^{3}}$ :
\begin{gather}
\left( x^{\prime }\right) _{(g),DSR4}^{\mu }(\left\{ x\right\}
_{m.},x^{5})=\left( x^{\mu }\right) _{(g),DSR4}^{\prime }(\left\{ x\right\}
_{m.},x^{5})=(1-\theta _{3}(g)S_{DSR4}^{3})_{~~\nu }^{\mu }(x^{5})x^{\nu
}\Leftrightarrow   \nonumber \\
\nonumber \\
\nonumber \\
\Leftrightarrow \left(
\begin{array}{c}
x_{(g),DSR4}^{0\prime }(x^{5}) \\
x_{(g),DSR4}^{1\prime }(x^{5}) \\
x_{(g),DSR4}^{2\prime }(x^{5}) \\
x_{(g),DSR4}^{3\prime }(x^{5})
\end{array}
\right) =\left(
\begin{array}{cccc}
1 & 0 & 0 & 0 \\
0 & 1 & \theta _{3}(g)b_{1}^{-2}(x^{5}) & 0 \\
0 & -\theta _{3}(g)b_{2}^{-2}(x^{5}) & 1 & 0 \\
0 & 0 & 0 & 1
\end{array}
\right) \left(
\begin{array}{c}
x^{0} \\
x^{1} \\
x^{2} \\
x^{3}
\end{array}
\right) =  \nonumber \\
\medskip   \nonumber \\
=\left(
\begin{array}{c}
x^{0} \\
x^{1}+\theta _{3}(g)b_{1}^{-2}(x^{5})x^{2} \\
-\theta _{3}(g)b_{2}^{-2}(x^{5})x^{1}+x^{2} \\
x^{3}
\end{array}
\right) \smallskip .
\end{gather}

{\bf 2} - {\em 3-d. deformed space-time (pseudo) rotations, or deformed
Lorentz boosts} (parameters $\underline{\zeta }(g)$ and generators $%
\underline{K_{DSR4}}(x^{5})$); they do {\em not} form a group (see Eq. (\ref
{Boosts}) below):

{\bf 2.1} - Infinitesimal boost with rapidity $\zeta _{1}(g)$ along $%
\widehat{x^{1}}$ :

\begin{gather}
\left( x^{\prime }\right) _{(g),DSR4}^{\mu }(\left\{ x\right\}
_{m.},x^{5})=\left( x^{\mu }\right) _{(g),DSR4}^{\prime }(\left\{ x\right\}
_{m.},x^{5})=(1-\zeta _{1}(g)K_{DSR4}^{1})_{~~\nu }^{\mu }(x^{5})x^{\nu
}\Leftrightarrow   \nonumber \\
\nonumber \\
\nonumber \\
\Leftrightarrow \left(
\begin{array}{c}
x_{(g),DSR4}^{0\prime }(x^{5}) \\
x_{(g),DSR4}^{1\prime }(x^{5}) \\
x_{(g),DSR4}^{2\prime }(x^{5}) \\
x_{(g),DSR4}^{3\prime }(x^{5})
\end{array}
\right) =\left(
\begin{array}{cccc}
1 & -\zeta _{1}(g)b_{0}^{-2}(x^{5}) & 0 & 0 \\
-\zeta _{1}(g)b_{1}^{-2}(x^{5}) & 1 & 0 & 0 \\
0 & 0 & 1 & 0 \\
0 & 0 & 0 & 1
\end{array}
\right) \left(
\begin{array}{c}
x^{0} \\
x^{1} \\
x^{2} \\
x^{3}
\end{array}
\right) =  \nonumber \\
\medskip   \nonumber \\
=\left(
\begin{array}{c}
x^{0}-\zeta _{1}(g)b_{0}^{-2}(x^{5})x^{1} \\
-\zeta _{1}(g)b_{1}^{-2}(x^{5})x^{0}+x^{1} \\
x^{2} \\
x^{3}
\end{array}
\right) \smallskip ;
\end{gather}
{\bf 2.2} - Infinitesimal boost with rapidity $\zeta _{2}(g)$ along $%
\widehat{x^{2}}$ :

\begin{gather}
\left( x^{\prime }\right) _{(g),DSR4}^{\mu }(\left\{ x\right\}
_{m.},x^{5})=\left( x^{\mu }\right) _{(g),DSR4}^{\prime }(\left\{ x\right\}
_{m.},x^{5})=(1-\zeta _{2}(g)K_{DSR4}^{2})_{~~\nu }^{\mu }(x^{5})x^{\nu
}\Leftrightarrow   \nonumber \\
\nonumber \\
\nonumber \\
\Leftrightarrow \left(
\begin{array}{c}
x_{(g),DSR4}^{0\prime }(x^{5}) \\
x_{(g),DSR4}^{1\prime }(x^{5}) \\
x_{(g),DSR4}^{2\prime }(x^{5}) \\
x_{(g),DSR4}^{3\prime }(x^{5})
\end{array}
\right) =\left(
\begin{array}{cccc}
1 & 0 & -\zeta _{2}(g)b_{0}^{-2}(x^{5}) & 0 \\
0 & 1 & 0 & 0 \\
-\zeta _{2}(g)b_{2}^{-2}(x^{5}) & 0 & 1 & 0 \\
0 & 0 & 0 & 1
\end{array}
\right) \left(
\begin{array}{c}
x^{0} \\
x^{1} \\
x^{2} \\
x^{3}
\end{array}
\right) =  \nonumber \\
\medskip   \nonumber \\
=\left(
\begin{array}{c}
x^{0}-\zeta _{2}(g)b_{0}^{-2}(x^{5})x^{2} \\
x^{1} \\
-\zeta _{2}(g)b_{2}^{-2}(x^{5})x^{0}+x^{2} \\
x^{3}
\end{array}
\right) \smallskip ;
\end{gather}
{\bf 2.3} - Infinitesimal boost with rapidity $\zeta _{3}(g)$ along $%
\widehat{x^{3}}$ :

\begin{gather}
\left( x^{\prime }\right) _{(g),DSR4}^{\mu }(\left\{ x\right\}
_{m.},x^{5})=\left( x^{\mu }\right) _{(g),DSR4}^{\prime }(\left\{ x\right\}
_{m.},x^{5})=(1-\zeta _{3}(g)K_{DSR4}^{3})_{~~\nu }^{\mu }(x^{5})x^{\nu
}\Leftrightarrow   \nonumber \\
\nonumber \\
\nonumber \\
\Leftrightarrow \left(
\begin{array}{c}
x_{(g),DSR4}^{0\prime }(x^{5}) \\
x_{(g),DSR4}^{1\prime }(x^{5}) \\
x_{(g),DSR4}^{2\prime }(x^{5}) \\
x_{(g),DSR4}^{3\prime }(x^{5})
\end{array}
\right) =\left(
\begin{array}{cccc}
1 & 0 & 0 & -\zeta _{3}(g)b_{0}^{-2}(x^{5}) \\
0 & 1 & 0 & 0 \\
0 & 0 & 1 & 0 \\
-\zeta _{3}(g)b_{3}^{-2}(x^{5}) & 0 & 0 & 1
\end{array}
\right) \left(
\begin{array}{c}
x^{0} \\
x^{1} \\
x^{2} \\
x^{3}
\end{array}
\right) =  \nonumber \\
\medskip   \nonumber \\
=\left(
\begin{array}{c}
x^{0}-\zeta _{3}(g)b_{0}^{-2}(x^{5})x^{3} \\
x^{1} \\
x^{2} \\
-\zeta _{3}(g)b_{3}^{-2}(x^{5})x^{0}+x^{3}
\end{array}
\right) .\smallskip
\end{gather}

The explicit form of the infinitesimal contravariant 4-vector $\delta
x_{(g),DSR4}^{\mu }(\left\{ x\right\} _{m.},x^{5})$, corresponding to an
element $g\in $ $SO(3,1)_{DEF.}$, is therefore:
\begin{gather}
\left\{
\begin{array}{c}
\delta x_{(g),DSR4}^{0}(\left\{ x\right\} _{m.},x^{5})=-\zeta
_{1}(g)b_{0}^{-2}(x^{5})x^{1}-\zeta _{2}(g)b_{0}^{-2}(x^{5})x^{2}-\zeta
_{3}(g)b_{0}^{-2}(x^{5})x^{3}=\medskip  \\
=b_{0}^{-2}(x^{5})(-\zeta _{1}(g)x^{1}-\zeta _{2}(g)x^{2}-\zeta
_{3}(g)x^{3})\bigskip , \\
\\
\delta x_{(g),DSR4}^{1}(\left\{ x\right\} _{m.},x^{5})=-\zeta
_{1}(g)b_{1}^{-2}(x^{5})x^{0}+\theta _{3}(g)b_{1}^{-2}(x^{5})x^{2}-\theta
_{2}(g)b_{1}^{-2}(x^{5})x^{3}=\medskip  \\
=-b_{1}^{-2}(x^{5})(\zeta _{1}(g)x^{0}-\theta _{3}(g)x^{2}+\theta
_{2}(g)x^{3})\bigskip , \\
\\
\delta x_{(g),DSR4}^{2}(\left\{ x\right\} _{m.},x^{5})=-\zeta
_{2}(g)b_{2}^{-2}(x^{5})x^{0}-\theta _{3}(g)b_{2}^{-2}(x^{5})x^{1}+\theta
_{1}(g)b_{2}^{-2}(x^{5})x^{3}=\medskip  \\
=-b_{2}^{-2}(x^{5})(\zeta _{2}(g)x^{0}+\theta _{3}(g)x^{1}-\theta
_{1}(g)x^{3}), \\
\bigskip  \\
\delta x_{(g),DSR4}^{3}(\left\{ x\right\} _{m.},x^{5})=-\zeta
_{3}(g)b_{3}^{-2}(x^{5})x^{0}+\theta _{2}(g)b_{3}^{-2}(x^{5})x^{1}-\theta
_{1}(g)b_{3}^{-2}(x^{5})x^{2}=\medskip  \\
=-b_{3}^{-2}(x^{5})(\zeta _{3}(g)x^{0}-\theta _{2}(g)x^{1}+\theta
_{1}(g)x^{2})\bigskip .
\end{array}
\right.   \nonumber \\
\end{gather}
The covariant components of such a 4-vector are
\begin{equation}
\left\{
\begin{array}{c}
\delta x_{0(g),DSR4}(\left\{ x\right\} _{m.})=-\zeta _{1}(g)x^{1}-\zeta
_{2}(g)x^{2}-\zeta _{3}(g)x^{3}\medskip , \\
\\
\delta x_{1(g),DSR4}(\left\{ x\right\} _{m.})=\zeta _{1}(g)x^{0}-\theta
_{3}(g)x^{2}+\theta _{2}(g)x^{3}\medskip , \\
\\
\delta x_{2(g),DSR4}(\left\{ x\right\} _{m.})=\zeta _{2}(g)x^{0}+\theta
_{3}(g)x^{1}-\theta _{1}(g)x^{3}\medskip , \\
\\
\delta x_{3(g),DSR4}(\left\{ x\right\} _{m.})=\zeta _{3}(g)x^{0}-\theta
_{2}(g)x^{1}+\theta _{1}(g)x^{2}\medskip .
\end{array}
\right.
\end{equation}
Comparing this last result with the expression (\ref{sol-flat}) of the
covariant Killing vector, we see the perfect matching between the space-time
rotational component of $\xi _{\mu }(\left\{ x\right\} _{m.})$ (unique for
all the 4-d. generalized Minkowski spaces) and the \ covariant 4-vector $%
\delta x_{\mu (g),DSR4}(\left\{ x\right\} _{m.})$ related to $SO(3,1)_{DEF.}$%
.

\subsection{The 4-d. deformed Lorentz algebra, i.e. the Lie algebra of the
4-d. deformed, homogeneous Lorentz group $SO(3,1)_{DEF.}$.}

Let us specialize Eq. (\ref{gen-Lor-1}) to the DSR4 case, in order to derive
the 4-d. deformed Lorentz algebra, i.e. the Lie algebra of the 4-d.
deformed, homogeneous Lorentz group $SO(3,1)_{DEF.}^{6\text{{\em par.}}}$.
We get:
\begin{gather}
\lbrack I_{DSR4}^{\alpha \beta }(x^{5}),I_{DSR4}^{\rho \sigma
}(x^{5})]=\medskip  \nonumber \\
\nonumber \\
\nonumber \\
=g_{DSR4}^{\alpha \sigma }(x^{5})I_{DSR4}^{\beta \rho
}(x^{5})+g_{DSR4}^{\beta \rho }(x^{5})I_{DSR4}^{\alpha \sigma
}(x^{5})+\medskip  \nonumber \\
\nonumber \\
-g_{DSR4}^{\alpha \rho }(x^{5})I_{DSR4}^{\beta \sigma
}(x^{5})-g_{DSR4}^{\beta \sigma }(x^{5})I_{DSR4}^{\alpha \rho
}(x^{5})=\medskip  \nonumber \\
\nonumber \\
\nonumber \\
=\delta ^{\alpha \sigma }\left( b_{0}^{-2}(x^{5})\delta ^{\alpha
0}-b_{1}^{-2}(x^{5})\delta ^{\alpha 1}-b_{2}^{-2}(x^{5})\delta ^{\alpha
2}-b_{3}^{-2}(x^{5})\delta ^{\alpha 3}\right) I_{DSR4}^{\beta \rho }(x^{5})+
\nonumber \\
\medskip  \nonumber \\
+\delta ^{\beta \rho }\left( \delta ^{\beta 0}b_{0}^{-2}(x^{5})-\delta
^{\beta 1}b_{1}^{-2}(x^{5})-\delta ^{\beta 2}b_{2}^{-2}(x^{5})-\delta
^{\beta 3}b_{3}^{-2}(x^{5})\right) I_{DSR4}^{\alpha \sigma }(x^{5})+
\nonumber \\
\medskip  \nonumber \\
-\delta ^{\alpha \rho }\left( \delta ^{\alpha 0}b_{0}^{-2}(x^{5})-\delta
^{\alpha 1}b_{1}^{-2}(x^{5})-\delta ^{\alpha 2}b_{2}^{-2}(x^{5})-\delta
^{\alpha 3}b_{3}^{-2}(x^{5})\right) I_{DSR4}^{\beta \sigma }(x^{5})+
\nonumber \\
\medskip  \nonumber \\
-\delta ^{\beta \sigma }\left( \delta ^{\beta 0}b_{0}^{-2}(x^{5})-\delta
^{\beta 1}b_{1}^{-2}(x^{5})-\delta ^{\beta 2}b_{2}^{-2}(x^{5})-\delta
^{\beta 3}b_{3}^{-2}(x^{5})\right) I_{DSR4}^{\alpha \rho }(x^{5})\smallskip .
\end{gather}

On account of the physical interpretation of the infinitesimal generators,
one has therefore the following kinds of commutation relations:

{\bf 1} - Commutator of generators of 3-d. deformed space rotations:
\begin{gather}
\lbrack I_{DSR4}^{ij}(x^{5}),I_{DSR4}^{lm}(x^{5})]=  \nonumber \\
\nonumber \\
\nonumber \\
=\delta ^{im}\left( \delta ^{i0}b_{0}^{-2}(x^{5})-\delta
^{i1}b_{1}^{-2}(x^{5})-\delta ^{i2}b_{2}^{-2}(x^{5})-\delta
^{i3}b_{3}^{-2}(x^{5})\right) I_{DSR4}^{jl}(x^{5})+  \nonumber \\
\medskip  \nonumber \\
+\delta ^{jl}\left( \delta ^{j0}b_{0}^{-2}(x^{5})-\delta
^{j1}b_{1}^{-2}(x^{5})-\delta ^{j2}b_{2}^{-2}(x^{5})-\delta
^{j3}b_{3}^{-2}(x^{5})\right) I_{DSR4}^{im}(x^{5})+  \nonumber \\
\medskip  \nonumber \\
-\delta ^{il}\left( \delta ^{i0}b_{0}^{-2}(x^{5})-\delta
^{i1}b_{1}^{-2}(x^{5})-\delta ^{i2}b_{2}^{-2}(x^{5})-\delta
^{i3}b_{3}^{-2}(x^{5})\right) I_{DSR4}^{jm}(x^{5})+  \nonumber \\
\nonumber \\
-\delta ^{jm}\left( \delta ^{j0}b_{0}^{-2}(x^{5})-\delta
^{j1}b_{1}^{-2}(x^{5})-\delta ^{j2}b_{2}^{-2}(x^{5})-\delta
^{j3}b_{3}^{-2}(x^{5})\right) I_{DSR4}^{il}(x^{5})=  \nonumber \\
\nonumber \\
\nonumber \\
\stackrel{\text{{\footnotesize ESC off on} }{\footnotesize i}\text{%
{\footnotesize \ , }}{\footnotesize j}}{=}-\delta
^{im}b_{i}^{-2}(x^{5})I_{DSR4}^{jl}(x^{5})-\delta
^{jl}b_{j}^{-2}(x^{5})I_{DSR4}^{im}(x^{5})+  \nonumber \\
\medskip  \nonumber \\
+\delta ^{il}b_{i}^{-2}(x^{5})I_{DSR4}^{jm}(x^{5})+\delta
^{jm}b_{j}^{-2}(x^{5})I_{DSR4}^{il}(x^{5})\smallskip ;  \label{Rotations}
\end{gather}

{\bf 2} - Commutator of generators of 3-d. deformed boosts:
\begin{gather}
\lbrack I_{DSR4}^{i0}(x^{5}),I_{DSR4}^{j0}(x^{5})]=  \nonumber \\
\nonumber \\
\medskip  \nonumber \\
=\delta ^{i0}\left( \delta ^{i0}b_{0}^{-2}(x^{5})-\delta
^{i1}b_{1}^{-2}(x^{5})-\delta ^{i2}b_{2}^{-2}(x^{5})-\delta
^{i3}b_{3}^{-2}(x^{5})\right) I_{DSR4}^{0j}(x^{5})+  \nonumber \\
\medskip  \nonumber \\
+\delta ^{0j}\left( \delta ^{j0}b_{0}^{-2}(x^{5})-\delta
^{j1}b_{1}^{-2}(x^{5})-\delta ^{j2}b_{2}^{-2}(x^{5})-\delta
^{j3}b_{3}^{-2}(x^{5})\right) I_{DSR4}^{i0}(x^{5})+  \nonumber \\
\medskip  \nonumber \\
-\delta ^{ij}\left( \delta ^{i0}b_{0}^{-2}(x^{5})-\delta
^{i1}b_{1}^{-2}(x^{5})-\delta ^{i2}b_{2}^{-2}(x^{5})-\delta
^{i3}b_{3}^{-2}(x^{5})\right) I_{DSR4}^{00}(x^{5})+  \nonumber \\
\medskip  \nonumber \\
-\delta ^{00}\left( \delta ^{00}b_{0}^{-2}(x^{5})-\delta
^{01}b_{1}^{-2}(x^{5})-\delta ^{02}b_{2}^{-2}(x^{5})-\delta
^{03}b_{3}^{-2}(x^{5})\right) I_{DSR4}^{ij}(x^{5})=  \nonumber \\
\nonumber \\
\medskip  \nonumber \\
=-b_{0}^{-2}(x^{5})I_{DSR4}^{ij}(x^{5});\medskip  \label{Boosts}
\end{gather}

{\bf 3} - ''Mixed'' commutator of 3-d. deformed space and boost generators:

\begin{gather}
\lbrack I_{DSR4}^{ij}(x^{5}),I_{DSR4}^{k0}(x^{5})]=  \nonumber \\
\nonumber \\
\medskip  \nonumber \\
=\delta ^{i0}\left( \delta ^{i0}b_{0}^{-2}(x^{5})-\delta
^{i1}b_{1}^{-2}(x^{5})-\delta ^{i2}b_{2}^{-2}(x^{5})-\delta
^{i3}b_{3}^{-2}(x^{5})\right) I_{DSR4}^{jk}(x^{5})+  \nonumber \\
\medskip  \nonumber \\
+\delta ^{jk}\left( \delta ^{j0}b_{0}^{-2}(x^{5})-\delta
^{j1}b_{1}^{-2}(x^{5})-\delta ^{j2}b_{2}^{-2}(x^{5})-\delta
^{j3}b_{3}^{-2}(x^{5})\right) I_{DSR4}^{i0}(x^{5})+  \nonumber \\
\medskip  \nonumber \\
-\delta ^{ik}\left( \delta ^{i0}b_{0}^{-2}(x^{5})-\delta
^{i1}b_{1}^{-2}(x^{5})-\delta ^{i2}b_{2}^{-2}(x^{5})-\delta
^{i3}b_{3}^{-2}(x^{5})\right) I_{DSR4}^{j0}(x^{5})+  \nonumber \\
\medskip  \nonumber \\
-\delta ^{j0}\left( \delta ^{j0}b_{0}^{-2}(x^{5})-\delta
^{j1}b_{1}^{-2}(x^{5})-\delta ^{j2}b_{2}^{-2}(x^{5})-\delta
^{j3}b_{3}^{-2}(x^{5})\right) I_{DSR4}^{ik}(x^{5})=  \nonumber \\
\nonumber \\
\medskip \medskip  \nonumber \\
\stackrel{\text{{\footnotesize \ ESC off on}{\small \ }}i\text{%
{\footnotesize \ and }}j\text{{\footnotesize , as before}}}{=}-\delta
^{jk}b_{j}^{-2}(x^{5})I_{DSR4}^{i0}(x^{5})+\delta
^{ik}b_{i}^{-2}(x^{5})I_{DSR4}^{j0}(x^{5})\smallskip .  \label{Mixed}
\end{gather}

In the ''self-representation'' basis of $SO(3,1)_{DEF.}^{6\text{{\em .}}}$,
it is easy to show that commutation relations (\ref{Rotations})-(\ref{Mixed}%
) read\footnote{%
Use has been made of the relation
\begin{gather*}
\epsilon _{ims}\epsilon _{jrs}b_{s}^{-2}(x^{5})= \\
\\
=\left( \delta _{ij}\delta _{mr}-\delta _{ir}\delta _{mj}\right) \left(
\sum_{k=1}^{3}(1-\delta _{ik})(1-\delta _{mk})b_{k}^{-2}(x^{5})\right)
\smallskip ,
\end{gather*}
which generalizes to the DSR4 case the well-known formula $\epsilon
_{ims}\epsilon _{jrs}=\delta _{ij}\delta _{mr}-\delta _{ir}\delta _{jm}$.}:
\begin{equation}
\left\{
\begin{array}{l}
i)\text{ \ }[S_{DSR4}^{i}(x^{5}),S_{DSR4}^{j}(x^{5})]\stackrel{\text{%
{\footnotesize ESC on}}}{=} \\
\\
\\
=\left( \sum_{s=1}^{3}(1-\delta _{is})((1-\delta
_{js})b_{s}^{-2}(x^{5})\right) \epsilon _{ijk}S_{DSR4}^{k}(x^{5})=\epsilon
_{ijk}b_{k}^{-2}(x^{5})S_{DSR4}^{k}(x^{5}), \\
\\
\medskip  \\
\\
ii)\text{ \ }[K_{DSR4}^{i}(x^{5}),K_{DSR4}^{j}(x^{5})]\stackrel{\text{%
{\footnotesize ESC on}}}{=}-b_{0}^{-2}(x^{5})\epsilon
_{ijk}S_{DSR4}^{k}(x^{5}), \\
\\
\medskip  \\
\\
iii)\text{ \ }[S_{DSR4}^{i}(x^{5}),K_{DSR4}^{j}(x^{5})]\stackrel{\text{%
{\footnotesize ESC on on }}l\text{{\footnotesize , ESC off on }}j}{=} \\
\\
\\
=\epsilon _{ijl}K_{DSR4}^{l}(x^{5})\left( \sum_{s=1}^{3}\delta
_{js}b_{s}^{-2}(x^{5})\right) \medskip =\epsilon
_{ijl}b_{j}^{-2}(x^{5})K_{DSR4}^{l}(x^{5}),
\end{array}
\right.   \label{Final-algebra}
\end{equation}
which define the deformed Lorentz algebra of generators of $SO(3,1)_{DEF.}^{6%
\text{{\em .}}}$.

\qquad Such relations generalize to the DSR4 case the infinitesimal
algebraic structure of the standard homogeneous Lorentz group $SO(3,1)$.
They admit interpretations completely analogous to those of the usual
Lorentz algebra:

$i)$ of Eq. (\ref{Final-algebra}) expresses the closed nature of the algebra
of the deformed rotation generators; consequently the 3-d. deformed space
rotations form a 3-parameter subgroup of $SO(3,1)_{DEF.}^{6\text{{\em .}}}$,
namely $SO(3)_{DEF.}$;

on the contrary, the deformed boost generator algebra is not closed
(according to $ii)$ of Eq. (\ref{Final-algebra}), and then the deformed
boosts do {\em not} form a subgroup of the deformed Lorentz group. This
implies that $SO(3,1)_{DEF.}^{6\text{{\em .}}}$ {\em cannot} be considered
the product of two its subgroups;

this is further confirmed by the non-commutativity of deformed space
rotations and boosts, expressed by $iii)$ of Eq. (\ref{Final-algebra}).

Moreover, $i)$ and $iii)$ of Eq. (\ref{Final-algebra}) show that both $%
\underline{S_{DSR4}}(x^{5})$ and $\underline{K_{DSR4}}(x^{5})$ behave as
3-vectors under deformed spatial rotations.

\section{Conclusions}

We want first to stress that, in the 4-dimensional case with the hyperbolic
metric signature $\left( S=3,T=1\right) $, in the limit
\begin{gather}
g_{\mu \nu ,DSR4}(x^{5})\rightarrow g_{\mu \nu ,SSR4}\Leftrightarrow
\nonumber \\
\nonumber \\
\nonumber \\
\Leftrightarrow \delta _{\mu \nu }(\delta _{\mu 0}b_{0}^{2}(x^{5})-\delta
_{\mu 1}b_{1}^{2}(x^{5})-\delta _{\mu 2}b_{2}^{2}(x^{5})-\delta _{\mu
3}b_{3}^{2}(x^{5}))\rightarrow  \nonumber \\
\medskip  \nonumber \\
\rightarrow \delta _{\mu \nu }(\delta _{\mu 0}-\delta _{\mu 1}-\delta _{\mu
2}-\delta _{\mu 3})\Leftrightarrow b_{\mu }^{2}(x^{5})\rightarrow 1,\text{ \
}\forall \mu =0,1,2,3\smallskip
\end{gather}
all results valid at group-transfomation level in DSR4 reduce to the
standard ones in SR.

Moreover, let us recall that the (parametric) dependence of the metric of a
generalized Minkowski space on the set $\left\{ x\right\} _{n.m.}$ of
non-metrical coordinates reflects itself also at the group level. In
particular, such a dependence shows up in:

1) the $N\times N$ matrix representation of the infinitesimal generators;

2) the infinitesimal group transformations;

3) the structure constants of the Lie algebra of generators .

Let us also notice that, since to {\em any} fixed value $\left\{ \bar{x}%
\right\} _{n.m.}$ of $\left\{ x\right\} _{n.m.}$ there corresponds a
generalized Minkowski space $\widetilde{M_{N}}(\left\{ \bar{x}\right\}
_{n.m.})$, we have a family of $N$-d. generalized Minkowski spaces
\begin{equation}
\left\{ \widetilde{M_{N}}(\left\{ x\right\} _{n.m.})\right\} _{\left\{
x\right\} _{n.m.}\in R_{\left\{ x\right\} _{n.m.}}},  \label{family}
\end{equation}
where $R_{\left\{ x\right\} _{n.m.}}$ is the range of the set $\left\{
x\right\} _{n.m.}$; if the cardinality of the range of each element of the
set $\left\{ x\right\} _{n.m.}$ is infinite, the cardinality of $R_{\left\{
x\right\} _{n.m.}}$ (and of the family (\ref{family})) is $\infty
^{N_{n.m.}} $. In correspondence, one gets a family of generalized
Poincar\'{e} groups
\begin{equation}
\left\{ P(S,T)_{GEN.}^{N(N+1)/2}(\left\{ x\right\} _{n.m.})\right\}
_{\left\{ x\right\} _{n.m.}\in R_{\left\{ x\right\} _{n.m.}}},
\end{equation}
with the same cardinality structure as (\ref{family}).

This can be summarized in the following scheme:

\begin{gather}
\left.
\begin{array}{c}
\text{{\em (Iper)spatial\ level} of }N\text{-d.\bigskip\ generalized
Minkowski spaces:} \\
\text{\bigskip } \\
1)\left\{ \widetilde{M_{N}}(\left\{ x\right\} _{n.m.})\right\} _{\left\{
x\right\} _{n.m.}\in R_{\left\{ x\right\} _{n.m.}}}\bigskip \\
\\
\\
2)\text{ }\widetilde{M_{N}}(\left\{ x\right\} _{n.m.})\equiv \widetilde{M_{N}%
}(\left\{ \bar{x}\right\} _{n.m.})\bigskip
\end{array}
\right\} \text{ }\Leftrightarrow \text{ \ \medskip \medskip \bigskip }
\nonumber \\
\nonumber \\
\\
\Leftrightarrow \left\{
\begin{array}{c}
\medskip \text{{\em Group level} of related maximal Killing groups:\bigskip }
\\
\\
1)\left\{ P(S,T)_{GEN.}^{N(N+1)/2}(\left\{ x\right\} _{n.m.})\right\}
_{\left\{ x\right\} _{n.m.}\in R_{\left\{ x\right\} _{n.m.}}}=\bigskip \\
\\
=\left\{ SO(T,S)_{GEN.}^{N(N-1)/2}(\left\{ x\right\} _{n.m.})\otimes
_{s}Tr.(T,S)_{GEN.}^{N\text{{\small \ }}}(\left\{ x\right\} _{n.m.})\right\}
_{\left\{ x\right\} _{n.m.}\in R_{\left\{ x\right\} _{n.m.}}}\bigskip \\
\\
\\
2)\text{ }P(S,T)_{GEN.}^{N(N+1)/2}(\left\{ x\right\} _{n.m.})\equiv
P(S,T)_{GEN.}^{N(N+1)/2}(\left\{ \bar{x}\right\} _{n.m.})\bigskip .
\end{array}
\right. \bigskip  \nonumber
\end{gather}

In the forthcoming papers, we will discuss the finite structure of the
space-time rotation groups, and the translation component of the maximal
Killing group of generalized Minkowski spaces.

\bigskip\

\ \ \ \ \ \ \ \ \ \ \ \ \

\bigskip {\bf References\bigskip }

\bigskip {\bf \lbrack 1]} \ F. Cardone and R. Mignani: ''On a nonlocal
relativistic kinematics'', INFN\medskip\ preprint n.910 (Roma, Nov. 1992);
{\it Grav. \& Cosm.} {\bf 4} , 311 (1998);\ {\it Found. Phys. }{\bf 29},
1735 (1999); \ {\it Ann. Fond. L. de Broglie }{\bf 25 }, 165 (2000).

\bigskip

{\bf [2]} \ F. Cardone, R. Mignani, and R.M. Santilli: {\it J. Phys.}G {\bf %
18}, L61, L141 (1992).\medskip

{\bf [3]} \ F. Cardone and R. Mignani: \ {\it JETP} {\bf 83}, 435 [{\it Zh.
Eksp. Teor. Fiz.} {\bf 110}, 793]\medskip\ (1996); F. Cardone, M. Gaspero,
and R. Mignani: {\it Eur. Phys. J.} C {\bf 4}, 705 (1998).\medskip

{\bf [4]} F.Cardone e R.Mignani : {\it Ann. Fond. L. de Broglie}, {\bf 23} ,
173 (1998);\ F. Cardone, R. Mignani, and V.S. Olkhovski: {\it J. de Phys.I
(France)}{\bf \ 7}, 1211\medskip\ (1997); {\it Modern Phys. Lett.} B {\bf 14}%
, 109 (2000).\medskip

{\bf [5]} \ F. Cardone and R. Mignani: {\it Int. J. Modern Phys.} A {\bf 14}%
, 3799 (1999).\medskip

{\bf [6]} \ F. Cardone, M. Francaviglia, and R. Mignani: {\it Gen. Rel. Grav.%
} {\bf 30}, 1619\medskip\ (1998); {\it ibidem}, {\bf 31}, 1049 (1999); {\it %
Found. Phys.\medskip\ Lett.} {\bf 12}, 281, 347 (1999){\it .}

{\bf [7] }A. Marrani: {\it ''Simmetrie di Killing di Spazi di Minkowski
generalizzati''} ({\it ''Killing Symmetries of Generalized Minkowski Spaces''%
}) (Laurea Thesis), Rome, October 2001 (in Italian).

{\bf [8]} F. Cardone, A. Marrani, and R. Mignani, {\it Found. Phys.} {\it %
Lett.} {\bf 16}, 163 (2003), {\tt hep-th/0505032}.\bigskip

\end{document}